%% file: main.tex
\documentclass[12pt]{article}
\usepackage[margin=1.25in]{geometry}
\usepackage{tikz}
\usetikzlibrary{patterns}
\usepackage[font=small,labelfont=bf]{caption}

\usepackage{amsmath, amssymb, amsthm}
\usepackage{mathtools}
\usepackage{natbib}
\usepackage[title]{appendix}
\usepackage{hyperref}
\hypersetup{
    colorlinks=true,
    linkcolor=black,
    filecolor=black,      
    citecolor=black,
    urlcolor=blue,
    }
\usepackage{multirow,array}
\usepackage{dirtytalk}
\usepackage{setspace}
\newtheorem{lemma}{Lemma}
\newtheorem{definition}{Definition}

\newtheorem{corollary}{Corollary}

\newtheorem{theorem}{Theorem}

\newcommand{\expdomain}{\mathcal{E}}

\newcommand{\comp}{\circ}

\newcommand{\supp}{\text{supp}}
\newcommand{\de}{\,d}
\newcommand{\id}{\mathbb{I}}
\newcommand{\sol}{\mathcal{S}}
\newcommand{\indicator}{\mathbf{1}}

\date{October 19, 2023\\
\href{https://grubbini.github.io/papers/BIC.pdf}{\large{Most recent version}}
}

\title{Mechanism Design without Rational Expectations \\ \Large{Job Market Paper}}
\author{Giacomo Rubbini\thanks{\scriptsize
{Department of Economics, Brown University, \href{mailto:giacomo_rubbini@brown.edu}{giacomo\_rubbini@brown.edu}. 
I am indebted to Roberto Serrano for his guidance and support. I wish to thank Pedro Dal Bò, Pietro Dall'Ara, Geoffroy De Clippel, Jack Fanning, Ricardo Fonseca, Takashi Kunimoto, Teddy Mekonnen, Zeky Murra Anton, Cosimo Petracchi, Marco Petterson,  Kareen Rozen, Rene Saran, Silvio Sorbera, Rajiv Vohra, and participants in various conferences and seminars for useful comments and suggestions. All errors are my own.
}}
}
\begin{document}

\pagenumbering{gobble}
\maketitle
\vspace{-0.9cm}
\begin{abstract}
\footnotesize{
\noindent Is incentive compatibility still necessary for implementation if we relax the rational expectations assumption? This paper proposes a generalized model of implementation that does not assume agents hold rational expectations and characterizes the class of solution concepts requiring Bayesian Incentive Compatibility (BIC) for full implementation. Surprisingly, for a broad class of solution concepts, full implementation of functions still requires BIC even if rational expectations do not hold. This finding implies that some classical results, such as the impossibility of efficient bilateral trade (Myerson \& Satterthwaite, 1983), hold for a broader range of non-equilibrium solution concepts, confirming their relevance even in boundedly rational setups.}
    \end{abstract}
    \vspace{0.15cm}
        \vspace{0.15cm}
    \newpage
\pagenumbering{arabic}

\section{Introduction}
Can a planner implement a given social goal by designing rules of interaction between agents when these agents hold private information they can exploit to their advantage? The answer depends on how this interaction pans out, and the literature on mechanism design and implementation has extensively explored this problem using a variety of game-theoretic solution concepts. 

While Bayesian Nash Equilibrium (BNE) remains a popular solution concept, insights from the experimental and behavioral literatures have highlighted that equilibrium models may not accurately predict agents' behavior in many settings. In these settings---for instance, when agents face a given interaction for the first time---the assumption that agents correctly anticipate their opponents' strategies (that is, that agents have rational expectations) feels particularly unpalatable. 

It remains unclear whether alternative solution concepts allow full implementation of a broader class of social choice rules than BNE. Recent results about full implementation of social choice functions (SCFs) in non-equilibrium solution concepts suggest that the answer may be negative. For instance, \cite{declippel.saran.serrano.2019} and \cite{kunimoto.saran.serrano.2020} prove that Bayesian Incentive Compatibility (BIC) is still necessary for full implementation of functions in level-\emph{k} reasoning and interim correlated rationalizability (ICR). In contrast, results are more permissive for full level-\emph{k} implementation of social choice sets (SCSs), for which BIC is no longer necessary \citep{declippel.saran.serrano.2019}. 

This paper studies the limits of full implementation by characterizing the class of all solution concepts such that BIC is necessary for full implementation. Our results suggest that we can generally not expect to dramatically expand the set of implementable SCFs by moving to non-equilibrium solution concepts, while results about SCSs are more permissive. 

Our novel approach turns on its head implementation theory's standard approach of fixing a solution concept and then deriving necessary conditions for full implementation, allowing us to search for a deeper property linking all solution concepts requiring BIC for full implementation. Other than providing useful guidance about the possibility of implementing non-BIC social choice rules, these results allow us to extend some classical findings in the literature (for example,   \citeauthor{myerson.satterthwaite.1983}'s \citeyearpar{myerson.satterthwaite.1983} impossibility theorem) to a large class of solution concepts.

To achieve this goal, we propose a generalized model of full implementation that allows agents to hold arbitrary expectations about their opponents’ strategies. This model allows us to encompass all solution concepts in which agents best respond to their (possibly heterogeneous) expectations about their opponents. Our model nests the ones in \cite{jackson.1991}, \cite{declippel.saran.serrano.2019}, \cite{kunimoto.saran.serrano.2020}, and \cite{kneeland.2022} as special cases, unifying previous results about the necessity of BIC for full implementation. 

For the case of implementation of SCFs, we show BIC is still necessary to implement functions if and only if the solution concept satisfies a novel property we call Weak Solution Consistency (WSC). This property can be interpreted as requiring that, for each type of each agent, there exists a solution of the mechanism such that she does not have any incentive to mimic a different type. Unlike regular incentive compatibility, WSC does not imply that this solution is the same for all types of all agents.\footnote{See discussion at the end of \autoref{model}.} Even if this property is not very restrictive, it is enough to establish the necessity of BIC, as full implementation of a function requires all the mechanism’s solutions to yield the outcome prescribed by the SCF.

Several solution concepts in the literature satisfy WSC---for instance, the level-\emph{k} model of \cite{declippel.saran.serrano.2019}, ICR \citep{kunimoto.saran.serrano.2020}, and BNE \citep{jackson.1991} satisfy this condition for any given mechanism. Notably, in the spirit of the so-called \textit{Wilson Doctrine},  WSC (and thus the necessity of BIC) does not hinge on the assumption of common knowledge of rationality. The epistemic argument in the Online Appendix shows that WSC is almost equivalent to requiring each type of each agent to know the type space and that  she can mimic another type by inducing a different solution of the mechanism. Both requirements feel natural, confirming that WSC is a mild restriction and that the class of solution concepts satisfying WSC is rather large.

By providing a characterization of the set of solution concepts that allow implementation of BIC SCFs, this paper also identifies which solution concepts allow for implementation of non-BIC functions. Cursed Equilibrium (CE; \citealt{eyster.rabin.2005}) and Naïve Bayesian Equilibrium (NBE; \citealt{gagnon-bartsch.pagnozzi.rosato.2021}) fall into the latter category, as they do not rule out the possibility that agents might not realize they could profitably mimic a different type. Existence of non-WSC solution concepts confirms that WSC has bite, and the characterization result hints at which solution concepts may be fruitful to investigate to study implementation of non-BIC SCFs.

\input{support_files/figure}

As for implementation of SCSs, WSC is not enough to establish the necessity of BIC for full implementation (\autoref{fig:diagram}).\footnote{The solution concept of \cite{declippel.saran.serrano.2019} is a case in point: even if their level-\emph{k} reasoning model satisfies WSC, they show in their Example 2 that it is possible to implement non-BIC SCSs.} The necessity of BIC for implementation of SCSs turns out to be close to assuming rational expectations. This, then, is  a relatively fragile result, unlikely to hold for most non-equilibrium models. WSC implies, however, that any implementable SCS must contain partially incentive compatible SCFs---that is, SCFs that provide only some types and agents with the right incentives not to misrepresent their private information. This last result confirms and extends the findings of \cite{kneeland.2022} for level-\emph{k} reasoning models.

The contrast between the results for implementation of SCFs and SCSs suggests that the necessity of BIC is mainly driven by the requirement that all solutions of the mechanism yield the same SCF when rational expectations do not hold. As agents understand that all solutions will lead to the same outcome in the case of implementation of SCFs, the same SCF must provide incentives to all agents not to misrepresent their type (that is, it must be BIC). If we allow different solutions to yield different outcomes instead (as in the case of full implementation of SCSs), each type may believe a different solution of the mechanism (and the associated SCF) will obtain. The planner no longer needs the same SCF to simultaneously incentivize all types of all agents, unless rational expectations hold. In a sense, decoupling agents' expectations allows the planner to decouple the incentives she provides them. As the rational expectations assumption makes this decoupling impossible, BIC is necessary for implementation in equilibrium solution concepts.

This discussion highlights a new tension that behavioral mechanism design faces: while having a unique outcome for all solutions offers starker predictions in applications, it  often delivers restrictive results regarding incentive compatibility. This tension is absent in equilibrium solution concepts: regardless of the number of solution outcomes, the rational expectations assumption ensures BIC is necessary for implementation. This result follows again from the fact that both the uniqueness requirement and the rational expectations assumption do not allow the planner to decouple the incentives she provides to each agent from the ones she provides to other agents. 

These results are important as they allow us to extend classical mechanism design findings to full implementation in any WSC solution concept. For the case of full implementation of functions, \autoref{applications} considers three applications that extend to all WSC solution concepts: the Revenue Equivalence Theorem \citep{myerson.1981}, the impossibility of ex-post efficient and budget-balanced bilateral trade \citep{myerson.satterthwaite.1983}, and the impossibility of full surplus extraction in auctions. These applications highlight that much of the underlying economic intuition for these results does indeed not hinge on the rational expectations assumption or the use of a particular equilibrium solution concept \textit{per se}, and it remains central for the case of boundedly rational agents as well.

This paper advances the literature on mechanism design and bounded rationality  by providing a methodology to answer open questions about implementation with and without rational expectations. Unlike previous works, this paper investigates the robustness of the necessity of BIC to changes in the solution concept. Previous papers focused instead on robustness in other model features---for example, \cite{saran.2011} characterizes the domain of preferences on which the revelation principle holds, while \cite{artemov.kunimoto.serrano.2013} prove that the restrictiveness of robust virtual implementation stems from a particular zero-measure set of beliefs. This work also relates to approaches considering a planner with an inaccurate model of agents' payoffs and beliefs, and in particular the literature about continuous \citep{oury.tercieux.2012, declippel.saran.serrano.2023} and robust \citep{bergemann.morris.2005} implementation. However, this paper focuses on a planner with an accurate model of payoffs and beliefs who is not sure how these map to the outcomes of strategic interaction, and it studies how sensitive restrictions on the set of implementable SCFs (such as BIC) are to changes in this mapping. 

Our approach is relevant to the study of different implementation frameworks or necessary conditions as well. For instance, \cite{declippel.saran.serrano.2023} show that imposing the continuity requirement on level-$k$ implementation does not make it significantly more restrictive as it does for BNE implementation. This result could be extended by characterizing the class of solution concepts for which it holds, highlighting what properties of the solution concept cause continuous implementation to impose significant additional restrictions on the class of implementable SCFs. 

\section{Model} \label{model}
The goal of the social planner is to select an alternative from a set $A$, conditional on some information privately held by the agents in set $I$. As usual in the literature, incomplete information is modeled by assuming that there exists a set of types $T_i$ for each agent $i\in I$ and that each agent knows her type but not the type of other players. Let $T= \times_{i \in I} T_i$ be the set of all possible type profiles. 

Agents' (interim) beliefs about the types of their opponents are denoted as $p_i: T_i \to \Delta(T_{-i})$---that is, when an agent is of type $t_i$, she believes other players are of types $t_{-i}$ with probability $p_i(t_{-i}|t_i)$.\footnote{For example, we can take $p_i(t_{-i}|t_i)$ to be the Bayesian posterior stemming from a common prior distribution $q: T \to (0,1)$ such that $q(T)=1$.} Assume also that for all $i \in I$ and $t \in T$, $p_i(\cdot|t_i)$ has full support.\footnote{This assumption is not necessary for the argument, but it makes the notation more convenient by avoiding stating results in terms of equivalent SCFs.} Preferences over lotteries have expected utility form, with Bernoulli utility $u_i: A \times T \to \mathbb{R}$. Abusing notation slightly, let $u_i(a, t)$ for $a \in \Delta(A)$ denote the utility agent $i$ derives from lottery $a$ when the type profile is $t$. 

The social planner seeks to implement a social choice function $f: T \to \Delta (A)$, and she does so by designing a mechanism $\gamma = (\mu, S)$, where $S= \times_{i \in I} S_i$ is an action space and $\mu: S \to \Delta(A)$ is an outcome function. Let $\Gamma$ denote the set of all possible mechanisms the planner can design. Once the planner has committed to a mechanism, agents choose a strategy profile $\sigma: T \to \Delta(S)$. We  denote the set of such functions as $\Sigma$. For all $i \in I$, we   let $\Sigma_i$ denote the set of all functions $\sigma_i: T_i \to \Delta(S_i)$ and  all functions $\Sigma_{-i}$ denote the set of $\sigma_{-i}:T_{-i} \to \Delta(S_{-i})$. For the rest of the paper, we  slightly abuse the notation above by considering $\mu(\sigma(t))$ to denote the lottery over $A$ induced by $\sigma(t)$ under the outcome function $\mu$. 

A key feature of rational expectations models is that agents' expectations turn out to be correct in equilibrium. For example, if $\sigma$ is a BNE, player $i$ expects her opponents to play $\sigma_{-i}$. To relax the rational expectations assumption, we  consider a more general theory of agents' expectations. For a given mechanism $\gamma$, let $e_{i, t_i} \in \Sigma_{-i}$ represent the expectations of type $t_i$ of agent $i$ about her opponent. The set of all possible expectations for mechanism $\gamma$ is denoted as $\expdomain(\gamma) = \times_{i \in I} \Sigma_{-i}$. As $e_{i, t_i}$ is a strategy profile for players $j \ne i$, we sometimes evaluate it at $t_{-i}$; thus, $e_{i, t_i}(t_{-i}) \in \Delta(S_{-i})$. To make the notation more compact, define a mapping $e_i: T_i \to \Sigma_{-i}$ that assigns $e_{i, t_i}$ to each type $t_i \in T_i$ and denote as $e$ any profile $(e_i)_{i \in I} \in \expdomain(\gamma)$. 

The formulation above implicitly assumes expectations are deterministic. However, given that we assume agents' preferences over lotteries admit an expected utility representation, this assumption does not cause further loss of generality. Agents are also allowed to expect their opponents' actions to be correlated as $e_{i, t_i} \in \Sigma_{-i}$, and we do not assume $\Sigma_{-i}$ have a product structure. This formulation makes it possible to accommodate models such as the ICR model of \cite{kunimoto.saran.serrano.2020}.\footnote{See \cite{dekel.fudenberg.morris.2007} for further discussion about the difference between independent and correlated interim rationalizability.}

Let   a \textit{theory of expectations} $E$ be any correspondence mapping each mechanism $\gamma$ to a subset $E(\gamma)$ of $\expdomain(\gamma)$. We  interpret $E(\gamma)$ as the expectations the model allows agents to hold. For example, ICR implicitly rules out the possibility that agents expect one of their opponents to play a dominated strategy (see  \autoref{wsc_examples} for some examples of models of expectations). As in \cite{declippel.saran.serrano.2019} and \cite{kunimoto.saran.serrano.2020}, we can interpret $E(\gamma)$ as the set of expectation profiles the planner believes could happen with nonzero probability. This interpretation is reflected in the implementation concept used below, which requires the outcome prescribed by $f$ to prevail regardless of the expectation profile considered. 

Define a \textit{theory of response} as any correspondence $R: E \times \gamma \to \Sigma$. A \textit{solution concept} is then a pair $\sol=(R, E)$ consisting of both a theory of expectation formation and a theory of how agents respond to these expectations. $\sol$ maps each mechanism $\gamma$ to a subset of $\Sigma$, which we can interpret as the mechanism's solutions.\footnote{To be precise, $\sol$ maps the game induced by mechanism $\gamma$ to a set of solutions. As the set of players, the type space, and the utility functions are taken as given, for the sake of brevity in the remainder of the paper let us  say $\sol$ associates each mechanism $\gamma$ with the set of its solutions $\sol(\gamma)$.} Formally, let $\sigma$ be a \textit{solution} to a mechanism $\gamma$ whenever $\sigma \in R(e)$ for $e \in E(\gamma)$. 

For all $\sigma_{-i} \in \Sigma_{-i}$, we  denote as $B_{i, t_i}(\sigma_{-i})$ the set of best replies for type $t_i$ of $i$ to the profile $\sigma_{-i}$.\footnote{The set of best responses should depend on the specific mechanism used as well, but we omit it to simplify notation.} That is, if $s_i \in B_{i, t_i}(\sigma_{-i})$, then for all $s_i' \in \Delta(S_i)$:
$$
\int_{T_{-i}} u_i(\mu(s_i, \sigma_{-i}(t_{-i})), t) \de p_i(t_{-i}|t_i) \geq \int_{T_{-i}} u_i(\mu(s_i', e_{i, t_i}(t_{-i})), t) \de p_i(t_{-i}|t_i). $$
Let  $B(e)$ denote the set of $\sigma \in \Sigma$ such that for all $i \in I$ and $t \in T$, $s=\sigma(t)$ is such that $s_i \in B_{i, t_{i}}(e_{i,t_i})$.

We say an SCF $f$ is \textit{fully implementable} whenever (1) there exists a mechanism $\gamma$ that has at least one solution  and   (2) every such solution yields the outcome prescribed by $f$. Formally,  an SCF $f$ is fully implementable in $\sol$ whenever there exists an implementing mechanism $\gamma$ such that $\mu(\sol(\gamma)) = f$ and  $R(e) \ne \emptyset$ for all $e \in E(\gamma)$. Moreover, let $\Gamma^f \subseteq \Gamma$ denote the class of all such mechanisms. 
Similarly,  an SCS $F \ne \emptyset$ is fully implementable if there exists $\gamma$ such that $\mu(\sol(\gamma)) = F$ and  $R(e) \ne \emptyset$ for all $e \in E(\gamma)$.  we denote the class of mechanisms implementing $F$ as $\Gamma^F\subseteq \Gamma$. 

For the remainder of the paper, we will refer to \textit{full implementation} simply as \textit{implementation} unless otherwise specified. We will moreover refer to the requirement that $\mu(\sol(\gamma)) = f$ as the \textit{uniqueness requirement}, as it demands that all solutions of the mechanism yield the very same SCF.

We say an SCF $f \in F$ is \textit{Bayesian Incentive Compatible} (BIC) for agent $i$ of type $t_i$ whenever, for all $t_i' \in T_i$:
$$
\int_{T_{-i}} u_i(f(t), t) \de p_i(t_{-i}|t_i) \geq 
\int_{T_{-i}} u_i(f(t_i',t_{-i}), t) \de p_i(t_{-i}|t_i).
$$
That is, type $t_i$ of agent $i$ has no incentive to pretend to be of a different type in the direct mechanism associated with the SCF. Moreover, let us say $f$ is BIC whenever it is BIC for all $i \in I$ and $t_i \in T_i$, and that an SCS $F$ is BIC whenever all $f \in F$ are BIC. 

Similarly, we say $f$ is \textit{Strict-if-Responsive Bayesian Incentive Compatible} (SIRBIC) for type $t_i$  of agent $i \in I$ whenever it is BIC for $t_i \in T_i$ and the inequality above is strict for all $t_i' \ne t_i$ such that $f(t_i', t_{-i}) \ne f(t)$ for some $t_{-i} \in T_{-i}$. Again,  $f$ is SIRBIC whenever it is SIRBIC for all types of all agents, and $F$ is SIRBIC if all SCF $f \in F$ are SIRBIC.

We derive most of our results about the necessity of BIC by imposing a mild requirement on the solution concept $\sol$. This requirement can be interpreted as requiring for each type $t_i$ of agent $i$ that there exists a solution of the mechanism such that she has no incentive to play the strategy associated with a different type.

\begin{definition}[Weak Solution Consistency (WSC)] \label{wsc_definition}
A solution concept $\sol$ satisfies WSC for a class of mechanisms $\tilde{\Gamma}\subseteq \Gamma$ whenever for all $\gamma \in \tilde{\Gamma}$ $i \in I$, $t_i \in T_i$ there exists $\sigma \in \sol(\gamma)$ such that for all $t_i' \in T_i$:
$$
\int_{T_{-i}} u_i (\mu(\sigma(t), t) \de p_i(t_{-i}|t_i) \geq  
\int_{T_{-i}} u_i (\mu(\sigma(t_i',t_{-i}), t) \de p_i(t_{-i}|t_i).
$$
We say $\sol$ satisfies WSC if it satisfies WSC for all $\gamma \in \Gamma$ such that $\sol(\gamma)\ne \emptyset$.
\end{definition}

Notice that this solution $\sigma$ need not be the same for all players $i$ as we do not require agents' expectations to be consistent anymore.\footnote{See also Remark 2 in \cite{kneeland.2022}.}. As we allow expectations to be type dependent, this solution need not be the same for any type $t_i$ of player $i$ either.\footnote{About this point, see the discussion of weak Interim Rationalizable Monotonicity in \cite{kunimoto.saran.serrano.2020}.} This  highlights that WSC is much weaker than incentive compatibility, which instead requires $\sigma$ to be the same for all types of all players.\footnote{See also the discussion about Total Weak Solution Consistency in \autoref{full_sets}.} Moreover,  \autoref{wsc_definition} directly implies that $\tilde{\sol}$ is WSC for $\tilde{\Gamma} \subseteq \Gamma$ whenever there exists a WSC solution concept $\sol$ such that $\sol(\gamma) \subseteq \tilde{\sol}(\gamma)$ for all $\gamma \in \tilde{\Gamma}$. 

Finally, we make a few additional technical assumptions. To make sure expected utility is well defined over the spaces discussed in the paper, let $A$, $T_i$, and $S_i$ be separable metrizable spaces endowed with the Borel sigma algebra; let product sets be endowed with the product topology; let the Bernoulli utility functions be bounded and continuous; and let SCF, mechanisms  and strategies be measurable functions.

\subsection{A Sufficient Condition for WSC}
More explicitly using the definitions of $E$ and $R$ from \autoref{model} enables us to provide a sufficient condition on $\sol$ for WSC that is both insightful and easy to check. 

We say a solution concept $\sol$ is \textit{Solution Consistent} (SC) for mechanism $\gamma$ whenever $R \subseteq B$ and for all $i \in I$ and $t_i \in T_i$ there exists $e \in E(\gamma)$ and $\sigma \in R(e)$ such that $(\sigma_i, e_{i, t_i}) \in \sol(\gamma)$.\footnote{As $(\sigma_i, e_{i, t_i}) \in \sol(\gamma)$ if and only if there exists $e' \in E(\gamma)$ such that $(\sigma_i, e_{i, t_i}) \in R(e')$, we can equivalently state that SC requires that for all $i \in I$ and $t_i \in T_i$ there exists $e, e' \in E(\gamma)$ and $\sigma \in R(e)$ such that $(\sigma_i, e_{i, t_i}) \in R(e')$.} 
We can see immediately that if $\sol$ satisfies SC for mechanism $\gamma$, then it satisfies WSC for the same mechanism. Let us moreover say $\sol$ is SC whenever it is SC for all $\gamma \in \Gamma$ such that $\sol(\gamma)\ne \emptyset$.

Solution Consistency is a rather mild requirement on $\sol$, as it requires only that agents best respond to what they expect from their opponents and that the resulting strategy profile could be justified as part of a solution to the mechanism. This means we can also interpret SC as demanding agents believe that their opponents display a minimal level of rationality: type $t_i$ of $i$ responds and expects her opponent to respond to some  expectation profile $e' \in E(\gamma)$---that is, $(\sigma_i, e_{i, t_i}) \in \sol(\gamma)$.

To make the interpretation above clearer,  consider an example in which SC does not hold. This is the case, for example, if we assume all players are either of level 1 or 0 in the models of \cite{declippel.saran.serrano.2019},  \cite{crawford.2021}, and \cite{kneeland.2022}. Suppose that, for all level-0 agents, the anchor is to play a dominated strategy. Then, profile $(\sigma_i(t_i'), \alpha_{-i}(t_{-i}))$ is not be a solution to the mechanism, generating an inconsistency between the profile  level-1 agents believe would prevail in the mechanism and what the solution concept actually is. Indeed, assuming all agents can be at least level 2 is crucial to ensure that SC holds for all $\gamma \in \Gamma$ (\autoref{level_k}). 

\section{A Bilateral Trading Example} \label{bilateral_example}
We can clarify the intuition about BIC's necessity for full implementation of functions by considering the example of bilateral trade between level-\emph{k} parties from \cite{crawford.2021} and the discussion of said example in \cite{declippel.saran.serrano.2019}.

Before moving to the example itself, we summarize how level-\emph{k} models of behavior work. Level-0 players of type $t_i$ are na\"ive and (non-strategically) play some anchor $\alpha_i(t_i)$, which is exogenous to the model. Level-1 agents instead believe their opponents to be level-0,  and so level-1 agents best respond to the belief their opponents are playing the anchor. We will say any such best response is a \textit{level-1 consistent} strategy, denoted as $\sigma^1$. For every level $k_i>1$, agents of level $k$ believe their opponents to be playing a \textit{level-$(k-1)$ consistent} strategy $\sigma^{k_i-1}$ and best respond accordingly. We say profile $\sigma$ is a solution to a game $\gamma$ whenever there exists a combination of levels $\{k_i\}_{i \in I}$ such that $k_i>0$ for all $i \in I$ and $\sigma_i$ is level-$k_i$ consistent for all $i \in I$.\footnote{As in \cite{declippel.saran.serrano.2019}, each agent's type describes only her beliefs about the payoff-relevant state: as levels do not affect preferences, they are not part of the description of an agent's type.} 

Suppose two risk-neutral parties trade an indivisible object with value $c$ for the seller and $v$ for the buyer, with both values distributed independently and uniformly between 0 and 1. They trade using as a protocol a $\frac{1}{2}$-double auction: the seller and the buyer respectively submit an ask $a$ and a bid $b$ for the object, and trade happens if and only if  $b \geq a$. In that case, the trading price is  $x = 0.5(a+b)$. The utility from not trading is 0 for both parties, while the utility from trading is $u_s = x - c$ and $u_b = v-x$ for the seller and the buyer, respectively.

As in \cite{crawford.2021}, we assume that the agents' anchor is uniformly distributed over $[0,1]$ and that both agents are of level $k=1$. Then there exists an SCF $f$ that is implementable but not BIC: the unique level-1 consistent strategies are to bid $\frac{2}{3}v$ for the buyer and to ask $\frac{2}{3}c +\frac{1}{3}$ for the seller, and the associated SCF stipulates that trade happens if and only if $2v \geq 2c +1$ at a price of $\frac{1}{6}(2v+2c+1)$. As remarked by \cite{declippel.saran.serrano.2019}, a buyer of value $v=0.5$ would then have an incentive to imitate a buyer of type $v=0.75$ to gain a positive payoff, violating BIC.

However, the same function is not implementable if the two agents could both be of level $k=2$. \cite{declippel.saran.serrano.2019} highlight that playing $\frac{2}{3}v+\frac{1}{9}$ for $v \geq \frac{1}{3}$ and $v$ otherwise is a best response for the buyer to the level-1 strategy of the seller. Similarly, playing $\frac{2}{3}c+\frac{2}{9}$ for $c \geq \frac{1}{3}$ and $c$ otherwise is a best reply for the seller to a level-1 buyer. The strategies form a solution to the mechanism considered, but the mechanism fails to implement $f$ as the two solutions lead to different outcomes.\footnote{As a matter of fact, it is straightforward to check that two level-2 players would trade for $v=c=\frac{1}{4}$, while two level-1 players would not trade for those values as $\frac{1}{2}=2v < 2c +1 = \frac{3}{2}$.}

This discrepancy follows because we need \textit{all} solutions of the mechanism to yield the same outcome for each type profile $t \in T$  for the mechanism to implement an SCF $f$. The argument, however, generalizes to any arbitrary mechanism $\gamma = (\mu, S)$.
Suppose $\mu$ has a solution $\sigma^1$ (so that $\sigma_i^1$ is a best reply to $\alpha_{-i}$ for all agents), and suppose such a solution induces a non-incentive compatible SCF. Then, $(\sigma_i^2, \sigma_{-i}^1)$ is a solution of the mechanism whenever $\sigma_i^2$ is a best reply to $\sigma_{-i}^1$, as player $i$ is best responding to level-1 consistent strategies while all other agents are best responding to their anchors. Moreover, it cannot be the case that  $\mu(\sigma^1) = \mu(\sigma^2)$. As $\sigma^1$ induces non-BIC $f$, there would then exist $i \in I$, $t_i, t_i' \in T_i$:
\begin{gather*}
    \int_{T_{-i}} u_i(
\mu(\sigma_i^2(t_i'), \sigma_{-i}^1(t_{-i})), t) \de p_i(t_{-i}|t_i) =\\
 \int_{T_{-i}} u_i(
\mu(\sigma_i^1(t_i'), \sigma_{-i}^1(t_{-i})), t) \de p_i(t_{-i}|t_i)>\\
 \int_{T_{-i}} u_i(
\mu(\sigma_i^1(t_i), \sigma_{-i}^1(t_{-i})), t) \de p_i(t_{-i}|t_i) =\\
\int_{T_{-i}} u_i(
\mu(\sigma_i^2(t_i), \sigma_{-i}^1(t_{-i})), t) \de p_i(t_{-i}|t_i)
\end{gather*}
Therefore, $\sigma^2$ is not a best reply to $\sigma^1$ for at least one type $t_i$ of player $i$. It must then be the case that $\mu(\sigma^1) \ne \mu(\sigma^2)$. This violates uniqueness, making it impossible for the mechanism to implement any non-incentive compatible SCF.\footnote{It would still be possible for the mechanism to implement a social choice \textit{set}. In fact, \cite{declippel.saran.serrano.2019} prove BIC is no longer necessary for level-\emph{k} implementation in this case.}

While the argument above relies on the properties of level-\emph{k} models (which are often solved recursively starting from the anchor), a similar result holds for a much larger class of solution concepts as well.
In particular, we prove any solution concept such that agents correctly predict the outcome of the mechanism makes BIC necessary for implementation. Equilibrium solution concepts are clearly part of this class, as agents correctly anticipate the strategies their opponents are using---that is, agents hold rational expectations. This class is broader, encompassing also solution concepts in which agents possibly hold heterogeneous and/or incorrect expectations about the strategies of their opponents. For instance, if we insist on full implementation of an SCF, the level-\emph{k} reasoning model of \cite{declippel.saran.serrano.2019}  and ICR \citep{kunimoto.saran.serrano.2020} fall into this class. 
\autoref{MS} discusses how, for full implementation of functions, the impossibility result of \cite{myerson.satterthwaite.1983} generalizes to this broader class of solution concepts, confirming its robustness even outside the rational expectations paradigm.

\section{Results} \label{incentive_compatibility}
We prove that BIC is still a necessary condition for implementation of functions if and only if the solution concept satisfies a novel property we called Weak Solution Consistency (\autoref{full_functions}). This property can be interpreted as requiring that for each type of each agent, there exists a solution to the mechanism in which she does not want to imitate a different type. WSC is satisfied by several solution concepts that have been considered in the literature, with some notable exceptions (\autoref{wsc_examples}). WSC is not enough, however, to establish that BIC is necessary for full implementation of sets (\autoref{full_sets}), for which  we need a condition stronger than WSC. 

\subsection{Full Implementation of Functions} \label{full_functions}
There is a tight link between WSC and the necessity of BIC for implementation of functions: BIC remains a necessary condition whenever the solution concept is WSC for all mechanisms implementing $f$ (and thus, whenever it is WSC for all mechanisms). Conversely, if $f$ is BIC, $\sol$ is WSC for the whole class of implementing mechanisms $\Gamma^f$.
\begin{theorem} \label{WSC_BIC_functions}
If $f$ is implementable in $\sol$ and $\sol$ is WSC for $\Gamma^f$, then it is BIC. If $f$ is BIC and implementable in $\sol$, then $\sol$ is WSC for $\Gamma^f$.
\end{theorem}
As WSC solution concepts allow each agent to pretend to be of a different type, any implementable SCF must provide agents an incentive not to misreport their type. The result then follows from the uniqueness requirement, which entails that the \textit{same} SCF must incentivize all agents not to mimic a different type. 

The full proof for the result is relegated to \autoref{proofs}. It is, however, instructive to discuss here a sketch of the argument for the ``if'' part to appreciate how WSC and the uniqueness requirement of full implementation drive the final result. The key step of the proof involves noticing that whenever   a solution $\sigma$ of mechanism $\gamma$ exists such that for type $t_i$ and all $t_i' \in T_i$:
$$
\int_{T_{-i}} u_i (\mu(\sigma(t), t) \de p_i(t_{-i}|t_i) \geq  
\int_{T_{-i}} u_i (\mu(\sigma(t_i',t_{-i}), t) \de p_i(t_{-i}|t_i). 
$$
Then, as any $f$ implemented by $\gamma$ is such that $\mu(\sigma)=f$ by the uniqueness requirement of full implementability, the inequality above yields:
$$
\int_{T_{-i}} u_i(f(t), t) \de p_i(t_{-i}|t_i) \geq 
\int_{T_{-i}} u_i(f(t_i',t_{-i}), t) \de p_i(t_{-i}|t_i).
$$
WSC ensures that such a solution $\sigma \in \sol$ exists for all $i \in I$ and $t_i \in T_i$: then, by the uniqueness requirement, if follows that all such solutions will yield $f$ as an outcome. This is enough to establish that $f$ is indeed BIC. 

 As argued below, the class of WSC solution concepts is rather broad, and it includes the level-\emph{k} model of \cite{declippel.saran.serrano.2019}, BNE, and ICR \citep{dekel.fudenberg.morris.2007}. For an example of a WSC solution concept not yet considered in the literature, see the discussion about $\Delta$-rationalizability in \autoref{rationalizability}.\footnote{Although the tools described in this paper can be used to investigate the necessity of BIC for implementation in other solution concepts as well, such an endeavor falls beyond this paper's scope.} 
\subsubsection{Necessity of SIRBIC}
It is also possible to use $E$ and $R$ to prove that the necessity of SIRBIC is a byproduct of the assumption that all best replies to an agent's expectations concur to form a solution to the mechanism, rather than to the use of a non-equilibrium solution concept. This is the case, for example, in \cite{declippel.saran.serrano.2019} and \cite{kunimoto.saran.serrano.2020}.
\begin{theorem}
\label{BIC_SIRBIC}
Suppose $f$ is BIC and implementable in $\sol$. If $R=B$, then $f$ is SIRBIC.
\end{theorem}
That is, if all best replies to a profile of expectations are solutions to the mechanism (as is the case for the examples discussed in \autoref{wsc_examples}), SIRBIC obtains for free from BIC and implementability.

\subsection{Full Implementation of Sets} \label{full_sets}
The results in \autoref{full_functions} suggest that the necessity of BIC is robust even if we consider non-equilibrium solution concepts for the case of full implementation of functions. This section considers implementation of social choice \textit{sets} instead.  \cite{declippel.saran.serrano.2019} and \cite{kneeland.2022}  prove that implementation of sets is more permissive than implementation of functions, as incentive compatibility of $F$ is not necessary for implementation. \autoref{WSC_BIC_sets} proves these positive results are due to the relaxation of the uniqueness requirement.
\begin{theorem} \label{WSC_BIC_sets}
If $F$ is implementable in WSC $\sol$, then for all $i \in I$ and $t_i \in T_i$, there exists $f^{i, t_i} \in F$ that is BIC for $i$ and $t_i$. Conversely, if $F$ is implementable and there exists $f^{i, t_i} \in F$ that is BIC for $i$ and $t_i$, then $\sol$ is WSC for $\Gamma^F$.
\end{theorem}

This result generalizes the standard incentive compatibility constraint, showing that only a form of \textit{partial} incentive compatibility is necessary for implementation of sets. Incentive constraints can be satisfied through a different function $f^{i, t_i}$ for each agent and type.\footnote{This point  is similar to the one \cite{kneeland.2022} makes about level-\emph{k} models.} A key implication is that the planner may be able to promise each type of each agent a different incentive $f^{i, t_i}$, exploiting heterogeneity in expectations across agents and types. Conversely, BIC requires the same function $f$ to satisfy the incentive constraints of all players and types, imposing $f^{i, t_i}=f^{j, t_j}$ for all $i,j \in I$, $t_i \in T_i$, and $t_j \in T_j$.  This provides intuition as to why implementation of sets is much more permissive than implementation of functions: as the planner is not restricted to a unique outcome for all solutions, she can decouple the incentives provided to each type of each player, possibly allowing for implementation of sets that do not contain any incentive compatible SCF.

We can also characterize the set of solution concepts that make BIC necessary for implementation when the uniqueness requirement is dropped. As the discussion of \autoref{WSC_BIC_sets} suggests, this class will feature concepts in which beliefs are consistent across players and types.

\begin{definition}[Total Weak Solution Consistency (TWSC)]
We say a solution concept $\sol$ is TWSC for mechanism $\gamma$ whenever for all $\sigma \in \sol$, $i \in I$, $t_i, t_i' \in T_i$:
$$
 \int_{T_{-i}} u_i(\mu(\sigma(t)), t) \de p_i(t_{-i}|t_i) \geq
     \int_{T_{-i}} u_i(\mu(\sigma_i(t_i'), \sigma_{-i}(t_{-i})), t) \de p_i(t_{-i}|t_i).
$$
\end{definition}

TWSC requires each solution of the mechanism to be incentive compatible, and it is almost equivalent to BNE. The only difference is that for all $i \in I$,  $\sigma_i(t_i)$ needs to be a better (rather than best) reply than $\sigma_i(t_i')$ to profile $\sigma_{-i}$. This comes very close to requiring rational expectations as well, as it entails that all types of all agents have consistent expectations about the outcome that will prevail in the mechanism.

\begin{theorem}\label{TWSC_BIC}
If $F$ is implementable in TWSC $\sol$, then it is BIC. If $F$ is implementable and BIC, then $\sol$ is TWSC for $\Gamma^F$.
\end{theorem}

Therefore, the necessity of BIC is a more \say{fragile} result whenever $F$ is not a singleton, as it stems from very restrictive assumptions about the solution concept. This fragility arises because a non-singleton $F$ allows different players to believe different outcomes will prevail in the mechanism. Therefore, it becomes no longer necessary for the \textit{same} outcome to simultaneously provides incentives not to misrepresent their private information to each type and agent.

That TWSC and WSC are equivalent for all mechanisms yielding the same outcome for all solutions confirms this intuition. It follows, then, that the BIC restriction on implementable social choice rules in the case of non-rational expectations can be imputed to the insistence on the uniqueness requirement---that is, on  insisting on fully implementing an SCF.  

\autoref{WSC_BIC_sets} provides a weaker result than the one obtained from \cite{kneeland.2022} for level-\emph{k} implementation. In this case, we can make use of $E$ in a more explicit way to bridge the gap between the two by requiring expectations  to not depend on each agent's type.\footnote{It would be possible to express the results above in terms of a modified WSC condition as well by substituting the qualifier \say{for all $i \in I$ and $t_i \in T_i$} with \say{for all $i \in I$.} However, stating the results in terms of expectations seems to be more intuitive. } \autoref{level_k} checks that this is indeed the case in \citeauthor{kneeland.2022}'s \citeyearpar{kneeland.2022} model. 
\begin{theorem}\label{constant_expectations}
If $F$ is implementable in SC $\sol$ and expectations are type-independent, for all $i \in I$ there exists $f \in F$ that is BIC for $i$.
\end{theorem}
In other words, \autoref{constant_expectations} tells us that if expectations are constant with respect to $i$'s type, then the same SCF $f$ must provide all types $t_i \in T_i$ with an incentive not to mimic another type.\footnote{For the result to go through, the argument in the proof of \autoref{constant_expectations} requires only that there exists one type-independent expectation in $E(\gamma)$ for each agent.} As for the difference between \autoref{WSC_BIC_functions} and \autoref{WSC_BIC_sets}, the comparison of \autoref{WSC_BIC_sets} and \autoref{constant_expectations} highlights how heterogeneity in expectations leads to a larger class of implementable social choice rules. Notice moreover \autoref{constant_expectations} entails any $f \in F$ is such that the payoff distribution it assigns to each agent depends on her type but not on her identity.\footnote{That is, whenever for all permutations $\pi$ of $I$, $i \in I$ and $t \in T$, we have $u_i(f(t), t) = u_{\pi(i)} (f(t_{\pi(1), ... , \pi(I)}), (t_{\pi(1), ... , \pi(I)}))$.} An example in this sense is the implementation of ex-post efficient allocations in auction problems with symmetric bidders and continuous distribution of values.

\section{Examples}
\label{wsc_examples}
As argued in the previous section, SC does not seem to be a particularly restrictive condition. It is indeed satisfied by various solution concepts proposed in the literature: BNE \citep{jackson.1991}, level-\emph{k} reasoning \citep{declippel.saran.serrano.2019, kneeland.2022}, and ICR \citep{kunimoto.saran.serrano.2020}. It is also satisfied by a weaker solution concept than BNE, one that  requires agents to share the same (not necessarily correct) expectations about their opponents.

WSC, in contrast, is not satisfied  by \citeauthor{eyster.rabin.2005}'s \citeyearpar{eyster.rabin.2005} Cursed Equilibrium model  and \citeauthor{gagnon-bartsch.pagnozzi.rosato.2021}'s \citeyearpar{gagnon-bartsch.pagnozzi.rosato.2021}  model of projection bias. This is because  in these models, the profile of strategies played by each agent as a response to her expectations and her expectations themselves generally do not form  a solution to the mechanism. These novel examples, together with \citeauthor{crawford.2021}'s \citeyearpar{crawford.2021} level-\emph{k} model with no level-2 agents, serve to confirm that WSC does indeed have bite.

\subsection{Level-\emph{k} Reasoning} \label{level_k}
The discussion in this section builds on the models of \cite{declippel.saran.serrano.2019} and \cite{kneeland.2022},  which assume that any profile of levels $k_i$ is possible, as long each level is lower than  an upper bound $\bar{k}\geq 2$---that is, $k_i \leq \bar{k}$ for all $i \in I$. 
% Unlike those models, however, our model does not assume agents' strategies are uncorrelated, allowing us to use our results  to slightly generalize their results about the necessity of BIC for full implementation.

Let $\alpha_{-i}: \Gamma \to \Sigma_{-i}$ be any correspondence assigning a profile of anchors to each mechanism $\gamma \in \Gamma$. For each agent $i \in I$, let $S_{-i}^{1}(\gamma|\alpha)$ denote the set of all {level-1 consistent} strategies $\sigma_{-i}: T_{-i} \to \Delta(S_{-i})$. Similarly, we  denote the set of best replies to \textit{level-($k_i-1$) consistent} strategy profiles as the set of \textit{level-$k_i$ consistent} strategies $S_{-i}^{k_i}(\gamma|\alpha)$.

We can now characterize the set of solutions for each mechanism $\gamma$ by setting $R=B$ and $E=E^{K, \alpha}$, where:
$$
E^{K, \alpha}(\gamma)= \{e \in \expdomain: e_{i,t_i} \in \{ \alpha_{-i}(\gamma) \} \cup \{\cup_{1 \leq k_i \leq \bar{k}} S_{-i}^{k_i-1}(\gamma|\alpha)\}, e_{i,t_i}=e_{i, t_i'}, \text{ for all }  i \in I, t_i, t_i' \in T_i\}.
$$

That is, the set of all $e \in \expdomain$ is such that each player $i$ expects the remaining players to play the anchor ($e_i \in \alpha_{-i}(\gamma)$) or to best respond as players of some level $k_i - 1$ ($e_i \in \cup_{1 \leq k_i \leq K} S_{-i}^{k_i-1}(\mu|\alpha)$). It is immediately apparent that any strategy profile such that each player's strategy is level-$k_i$ consistent for $k_i \geq 1$ is a solution of the mechanism.\footnote{In this case, expectations are type independent, allowing me to derive a slightly stronger result for full implementation of SCSs (\autoref{constant_expectations}).} 

This solution concept satisfies SC for all $\gamma \in \Gamma$. Because $\bar{k} \geq 2$, $E^{K, \alpha}(\gamma)$ contains at least one $e$ such that $e_i \in S_{-i}^{1}(\gamma|\alpha)$. Consider, then, that for all $i \in I$ and $t_i \in T_i$, any $\sigma \in B(e)=R(e)$. It is  clear, then, that $(\sigma_i, e_i) \in \sol(\gamma)$ because $\sigma_i$ is a level-2 consistent strategy and $e_i$ is a profile of level-1 consistent strategies.

The assumption that $\bar{k}\geq 2$ \citep{declippel.saran.serrano.2019, kneeland.2022} is  useful for excluding pathological cases in which a player expects her opponents just to play their anchor. This seems to explain why the findings of \cite{declippel.saran.serrano.2019} and \cite{kneeland.2022} differ from thos of  \cite{crawford.2021}, which instead proves it is possible to implement non-BIC SCFs. This possibility result  arises because \cite{crawford.2021} considers a setup with no level-2 players, allowing for the possibility that SC does not hold. 

\subsection{Interim Correlated and $\Delta$-Rationalizability}
\label{rationalizability}
\cite{kunimoto.saran.serrano.2020} study implementation using Interim Correlated Rationalizability (ICR) as a solution concept, finding that SIRBIC is a necessary condition for implementing SCFs.
 
Let $C=(C_i)_{i \in I}$ be a correspondence profile such that $C_i: T_i \to 2^{S_i}$ for all $i \in I$. Consider now the operator $b=(b_i)_{i \in I}$ iteratively eliminating strategies that are never a best response:
$$
b_i(C)[t_i] \equiv \left\{s_i \in S_i:
        \begin{array}{l}
        \mbox{$\exists \lambda_{i} \in \Sigma_{-i}$ such that: } \\
        (1)\ \supp (\lambda_{i}(t_{-i}))\subseteq C_{-i}(t_{-i}); \\
        (2)\ s_i \in \arg \max_{s_i'} \int_{T_{-i}} u_i(\mu(s_i', \sigma_{-i}(t_{-i})), t) \de p_i(t_{-i}|t_i) 
        \end{array}\right\}
$$
As argued in \cite{kunimoto.saran.serrano.2020}, by Tarski's theorem, there exists a largest fixed point of $b$, which is denoted as $C^{\gamma(T)}$. The authors then require that, for $f$ to be implementable,  there must exist a mechanism such that (1) the desired outcome obtains for all rationalizable strategy profiles and (2)  each type $t_i$ has at least one rationalizable action. 

We can then show that the class of ICR strategy profiles can be characterized by the following pair $(E^{ICR}, R^{ICR}):$\footnote{The requirement that $\supp(\sigma(t))|=1$ for all $t \in T$ arises because, as in \cite{kunimoto.saran.serrano.2020}, we focus on \textit{pure} strategies.}
$$
E^{ICR}(\gamma) = \{e \in \expdomain: \supp(e_{i, t_i}(t_{-i})) \subseteq C_{-i}^{\gamma(T)}(t_{-i}) \}
$$
$$
R^{ICR}(e) = \{ \sigma \in \Sigma: \sigma \in B(e), |\supp(\sigma(t))|=1 \text{ for all } t \in T \}
$$
This follows because $\sol^{ICR}(\gamma) = R^{ICR}(E^{ICR}) =C^{\gamma(T)}$ for all $\gamma \in \Gamma$. In fact, $\sigma \in B(e)$ for $e \in E(\gamma)$ implies that the unique profile $s$ in $\sigma$'s support is a rationalizable profile of actions, and thus it implies that $\sigma \in C^{\gamma(T)}$ and  $\sol \subseteq C^{\gamma(T)}$. Conversely, if $\sigma \in C^{\gamma(T)}$ for all $i \in I$ and $t_i \in T_i$, then there exists a $\lambda_i \in \Sigma_{-i}$ to which  $\sigma_i(t_i)$ is a best reply. Setting $e_{i, t_i} = \lambda_i$, then, is enough to achieve  $ C^{\gamma(T)}\subseteq \sol $.

ICR satisfies SC for a large class of mechanisms $\gamma \in \Gamma$---in particular, those in which $B_i(e_{i, t_i}) \ne \emptyset$ for all $i \in I$ and $t_i \in T_i$. This is the case, for example, if $A$ is finite as in \cite{kunimoto.saran.serrano.2020}. Then, for  any solution $\sigma \in \sol^{ICR}$ and $\tilde{\sigma}_i \in B_i(\sigma_{-i})$,  we have $(\tilde{\sigma}_i, \sigma_{-i}) \in \sol^{ICR}(\gamma)$ because $\sigma \in \sol^{ICR}(\gamma)$ entails that $\sigma_{-i}$ is rationalizable for all agents $j \ne i$ and that $\tilde{\sigma}_i$ is rationalized by the belief that $i$'s opponents are playing $\sigma_{-i}$.

The same argument applies even if we require, similarly to $\Delta$-rationalizability \citep{battigalli.siniscalchi.2003}, that agents' beliefs about their opponents' strategies lie in a pre-specified set. For each $i \in I$ and $t_i \in T_i$, let $\Delta^i$ map each mechanism $\Gamma$ to a set of allowed beliefs $\Delta_{i,t_i}(\gamma)$. Let $\Delta = (\Delta^i)_{i \in I}$. We can then redefine the operator $b$ as follows:
$$
b_i(C)[t_i] \equiv \left\{s_i \in S_i :
        \begin{array}{l}
        \mbox{$\exists \lambda_{i} \in \Delta^i \subseteq \Sigma_{-i}$ such that: } \\
        (1)\ \supp (\lambda_{i}(t_{-i}))\subseteq C_{-i}(t_{-i}); \\
        (2)\ s_i \in \arg \max_{s_i'} \int_{T_{-i}} u_i(\mu(s_i', \sigma_{-i}(t_{-i})), t) \de p_i(t_{-i}|t_i) 
        \end{array}\right\}
$$
Again, as $b$ is a monotone operator, Tarski's theorem implies that there exists a largest fixed point, which we again denote as $C^{\gamma(T), \Delta}$. The same argument as above then shows that the class of $\Delta$-rationalizable strategies $\sol^\Delta$ can be characterized through the following pair:
$$
E^{\Delta}(\gamma) = \{e \in \expdomain: \supp(e_{i, t_i}(t_{-i})) \subseteq C_{-i}^{\gamma(T), \Delta}(t_{-i}) \}
$$
$$
R^{\Delta}(e) = \{ \sigma \in \Sigma: \sigma \in B(e), |\supp(\sigma(t))|=1 \text{ for all } t \in T \}
$$

$\Delta$-rationalizable full implementation has not been considered in the literature, so it is not yet known whether BIC is necessary for full implementation.\footnote{\cite{artemov.kunimoto.serrano.2013} use $\Delta$-rationalizability as a solution concept. Differently from this paper, they study \textit{robust} virtual implementation by imposing restrictions on the set of beliefs agents may have about their opponents' types.} There is no obvious relation between the set of $\Delta$-rationalizable profiles and equilibrium profiles either, as the relation depends on the restrictions imposed by $\Delta$. For a simple example, focus on the following complete-information game:\footnote{We can think of it as a Bayesian game in which each player has only one type.}
\begin{table}[h!]
    \centering
\setlength{\extrarowheight}{2pt}
\begin{tabular}{cc|c|c|c|}
  & \multicolumn{1}{c}{} & \multicolumn{3}{c}{Player $C$} \\
  & \multicolumn{1}{c}{} & \multicolumn{1}{c}{$A$}  & \multicolumn{1}{c}{$B$}  & \multicolumn{1}{c}{$C$} \\\cline{3-5}
            & $A$ & $(2,2)$ & $(-2,-2)$ & $(-2,-2)$ \\ \cline{3-5}
Player $R$  & $B$ & $(-2,-2)$ & $(1,-1)$ & $(-1,1)$ \\\cline{3-5}
            & $C$ & $(-2,-2)$ & $(-1,1)$ & $(1,-1)$ \\\cline{3-5}
\end{tabular}
\end{table}
\\
This game admits only one pure-strategy equilibrium in which both players play $A$.\footnote{We focus on pure equilibria to keep our results comparable with those for Interim Correlated Rationalizability discussed above.} If $\Delta$ imposes no restriction on players' beliefs,  any pure-strategy profile is $\Delta$-rationalizable; therefore, any equilibrium is $\Delta$-rationalizable as well. Suppose now we restrict agents' beliefs to assign positive probability to $B$ and $C$ only. As $A$ is dominated when the opponent never plays $A$, any profile in which $A$ is played is now not $\Delta$-rationalizable, implying that the set of $\Delta$-rationalizable profiles is disjoint from the set equilibrium profiles.

We can then use \autoref{WSC_BIC_functions} and \autoref{WSC_BIC_sets} to derive a novel result about the necessity of BIC for $\Delta$-rationalizable implementation by proving that $\sol^{\Delta}$ is WSC whenever $\Delta$ satisfies the mild condition that it does not rule out that agents believe that their opponents will play a pure strategy. That is, for all $i \in I$, $\Delta^i$ contains a $\lambda_i$ whose support is a singleton for all $t_{-i} \in T_{-i}$. Let us again consider the class of mechanisms such that $B_i(e_{i, t_i}) \ne \emptyset$ for all $i \in I$ and $t_i \in T_i$. Because for each $i \in I$ and $t_i \in T_i$ there exists  a $\lambda_i \in \Delta^i$ whose support is a singleton for all $t_{-i} \in T_{-i}$, let us construct $\sigma^{i, t_i}=(\sigma_i, \lambda_i)$, where $\sigma_i \in B_i(\lambda_i)$. As $\sigma^{i, t_i} \in \sol^{\Delta}$, $\sol^{\Delta}$ is WSC; in fact, $\sigma_i^{i, t_i}$ is rationalized by $\lambda_i \in \Delta^i$, and $\sigma_j^{i, t_i}$ is rationalized by some belief in $\Delta^j$ for all $j \ne i$. 

\subsection{Bayesian Nash Equilibrium and Refinements}
The setup proposed in this paper can capture Bayesian Nash equilibrium if we impose the following:
$$
E^{BN}(\gamma) = \{ e \in \expdomain(\gamma): \exists \ \sigma \in \times_{i \in I} \Sigma_i \text{ s.t. } e_{i,t_i}= \sigma_{-i} \text{ for all } i \in I, t_i \in T_i, \sigma \in B((\sigma_{-i})_{i \in I})\}
$$
$$
R^{BN}(e) = \{ \sigma \in B(e): \sigma_{-i}=e_i \text{ for all } i \in I \}
$$
It is  clear then that the set of BNEs is equal to $R^{BN}(E^{BN}(\gamma))=\sol^BN(\gamma)$. In fact, if $\sigma \in \sol^BN(\gamma)$, then $\sigma_i \in B(\sigma_{-i})$ for all $i \in I$. On the other hand, if $\sigma$ is a BNE, it is immediate to notice that $(\sigma_{-i})_{i\in I} \in E^{BN}(\gamma)$ and thus $\sigma \in R^{BN}(e)$. Moreover, as long as $E^{BN}(\gamma) \ne \emptyset$, $R^{BN}(E^{BN}(\gamma)) \ne \emptyset$ as well. In fact, $R^{BN}$ just selects, among all profiles of best responses, the one satisfying rational expectations. $\sol^{BN}$ also satisfies SC for all $\gamma \in \Gamma$ because for all $i \in I$ and $t_i \in T_i$, the expectation profile $e' = (\sigma_{-i})_{i \in I}$ is such that $(\sigma_i, e_{i, t_i}) = \sigma \in B(e')$. The same argument applies to refinements of BNE as well (as undominated BNE), because they all satisfy the rational expectations assumption.

\subsection{Cursed Equilibrium}
This setup can also capture the \emph{Cursed Equilibrium} solution concept from \cite{eyster.rabin.2005}. This leads to the following theory of behavior:
$$
E^{CE}(\gamma) = \{e \in \expdomain(\gamma): \exists \ \sigma \in \times_{i \in I} \Sigma_i \text{ s.t. } e_{i,t_i}= \sigma_{-i} \text{ for all } i \in I, t_i \in T_i, \sigma \in B((\bar{\sigma}_{-i})_{i \in I})\}
$$
$$
R^{CE}(e) = \{\sigma \in B(e): \bar{\sigma}_{-i} = e_i  \},
$$
where:
$$
\Bar{\sigma}_{-i}(t_i) = \int_{T_{-i}} \sigma_{-i}(t_{-i}) \de p_i(t_{-i}|t_i).
$$
It is possible to prove that this solution concept is WSC for all mechanisms $\gamma$ if agents have private values  but  not otherwise. Intuitively, the reason  is that the payoff distribution that agents expect to achieve differs from the one the mechanism actually implements. For example, a fully cursed ($\chi=1$) agent $i$ expects the payoff from playing the action associated with type $t_i'$ to be $\int_{T_{-i}} u_i(f(t_i', \Bar{\id}_{-i} ), t) \de p_i(t_{-i}|t_i)$ rather than $\int_{T_{-i}} u_i(f(t_i', t_{-i} ), t) \de p_i(t_{-i}|t_i)$, where $\id$ represents the identity function.\footnote{The difference between these two expressions is immaterial for private-value auctions  but  not for common-value ones.}

To prove the result for private values, suppose $\sigma \in R^{CE}(E^{CE}(\gamma))$. Therefore, we have that for all $t_i \in T_i$ and $s_i \in \Delta(S_i)$:
\begin{gather*}
    (1-\chi) \int_{T_{-i}} u_i((\sigma(t_i), \sigma_{-i}(t_i)), t_i) \de p_i(t_{-i}|t_i) + \chi \int_{T_{-i}} u_i((\sigma(t_i), \Bar{\sigma}_{-i}(t_i)), t_i) \de p_i(t_{-i}|t_i) \geq \\
    (1-\chi) \int_{T_{-i}} u_i(s_i, \sigma_{-i}(t_i)), t_i) \de p_i(t_{-i}|t_i) + \chi \int_{T_{-i}} u_i(s_i, \Bar{\sigma}_{-i}(t_i)), t_i) \de p_i(t_{-i}|t_i).
\end{gather*}
Then, as $u_i(\cdot)$ does not depend on $t_{-i}$, by linearity of expected utility it follows that:
$$
\int_{T_{-i}} u_i((\sigma(t), t_i) \de p_i(t_{-i}|t_i) \geq \int_{T_{-i}} u_i(s_i, \sigma_{-i}(t_{-i}), t_i) \de p_i(t_{-i}|t_i).
$$
As $\sigma$ is a solution, this concludes the proof.

However, the same is not valid for all mechanisms $\gamma$ if an agent's payoff depends on the type of her opponents. This is because the profile $(\sigma_i, \bar{\sigma}_{-i})$ is typically not a solution to the mechanism, violating SC. To prove WSC is violated as well for any $\chi \in (0,1]$, we can construct the following two-player game: 

\begin{table}[h!]
        \centering
    \begin{tabular}{cc|c|c|}
      & \multicolumn{1}{c}{} & \multicolumn{2}{c}{Player $C$}\\
      & \multicolumn{1}{c}{} & \multicolumn{1}{c}{$A$}  & \multicolumn{1}{c}{$B$} \\\cline{3-4}
      \multirow{2}*{Player $R$}  & $A$ & $t_R,t_C$ & $t_R+\zeta t_C, 0$ \\\cline{3-4}
      & $B$ & $0,t_C+\zeta t_R$ & $0,0$ \\\cline{3-4}
    \end{tabular}
\end{table}
Where $t_i \in \{-1, 1\}$ for $i \in \{R, C\}$, each type profile happens with equal probability, and $\zeta \in (2, \frac{2}{1-\chi}).$\footnote{In the discussion below, the argument  focuses on the case of $\chi<1$. The case of $\chi=1$ follows from the same steps as long as $\zeta>2$.} 
The only Cursed Equilibrium of this game is for type $1$ to play $A$ and for type $-1$ to play $B$. To prove this, consider any solution $\sigma$ of the game. Then:
$$
\Bar{\sigma}_i(t_i)[A] = \frac{1}{2}\sigma_j(1)[A] +  \frac{1}{2}\sigma_j(-1)[A].
$$
The payoff of $B$ is always 0 for either player, while the payoff from playing $A$ is:
$$
t_i - \frac{1}{2}(1-\chi)\zeta ( \sigma_{-i}(1)[A] -\sigma_{-i}(-1)[A]).
$$
Type $1$ will play $A$ with probability 1 as long as:
$$
1 - \frac{1}{2}(1-\chi)\zeta >0 \iff \zeta < \frac{2}{1-\chi}.
$$
And type $t_i=-1$ will play $B$ with probability 1 whenever:
$$
-1 - \frac{1}{2}(1-\chi)\zeta <0 \iff \zeta > \frac{-2}{1-\chi}.
$$
There is therefore a pure Cursed Equilibrium in which both agents play $A$ if their type is $t_i=1$ and $B$ otherwise. Moreover, this is the unique Cursed Equilibrium of the game, and it does not satisfy WSC. In fact, for type $t_i=1$ of player $i$, $\zeta>2$ implies that:
$$
\frac{1}{2}t_i+\frac{1}{2}(t_i-\zeta) = 1-\frac{1}{2}\zeta<0.
$$
Thus, type $t_i=1$ would  like to mimic type $t_i=-1$ if she was not ignoring the correlation between her opponents' strategies and types: Cursed Equilibrium then allows for the implementation of non-BIC SCFs. 
\subsection{Naïve Bayesian Equilibrium}
\cite{gagnon-bartsch.pagnozzi.rosato.2021} propose a model of taste projection---that is, the tendency of agents to believe that their opponents' valuations of an object are more similar to their own than they actually are. The associated solution concept, Naïve Bayesian Equilibrium (NBE),  captures the idea that agents play according to the BNE of a \textit{perceived} game in which beliefs are distorted by taste projection.

To keep the discussion simple, let agents' values be private and independent.\footnote{More general versions of this model can be accommodated in the framework presented in the Online Appendix.} Let  $BNE(\gamma, \hat{\tau}(t_i))$ then denote the set of pure-strategy BNEs of mechanism $\gamma$ when agents believe that their opponents' values  are determined according to the random variable $\hat{\tau}(t_i)=\chi t_i +(1-\chi) \tau$ rather than the true random variable $\tau$. We can then define $\sol^{NBE}=(E^{NBE}, R^{NBE})$ as follows:
$$
E^{NBE}(\gamma) = \{ e \in \expdomain(\gamma): e_{i,t_i}= \sigma_{-i} \text{ for all } i \in I, t_i \in T_i, \text{ where } \sigma \in BNE(\gamma, t_i) \}
$$
$$
R^{NBE}(e) = \{ \sigma \in \Sigma: (\sigma_i, e_i) \in BNE(\gamma, \cdot) \text{ for all } i \in I\} 
$$

Let us say that $(A, T, I)$ is an \textit{economic environment} whenever, for all $t \in T$, $i,j \in I$ and $a \in A$, there exist $b, c \in A$ such that $u_i(b,t) > u_i(a,t)$ and $u_j(c,t)>u_j(a,t)$. Say also $(A, T, I)$ satisfies \textit{single crossing} whenever $u_i(a, t) \geq u_i(b,t)$ and  $u_i(a, t') \geq u_i(b,t')$ for all $i \in I$  imply that there exists a $c \in A$ such that $u_i(a,t) \geq u_i(c,t)$ and $u_i(c,t) > u_i(a,t)$.

\autoref{NBE_example} shows that NBE is generally not WSC.
\begin{theorem}\label{NBE_example}
    Let $|I| \geq 3$. If $\chi=1$ and $(A,T,I)$ satisfies the single-crossing and economic-environment assumptions, any SCF $f$ can be implemented in NBE via the Maskin mechanism \citep{maskin.1999}.
\end{theorem}
For $\chi =1$, the perceived mechanism is a complete-information game. Then, if $(A,I,T)$ satisfies the economic-environment and single-crossing assumptions, any SCF $f$ is implementable via the canonical mechanism in \cite{maskin.1999}, as $f$ vacuously satisfies  no-veto-power and Maskin-monotonicity. Moreover, the NBE is unique, as all BNEs of the Maskin mechanism prescribe the same action for agent $i$ of type $t_i$. Under these restrictions on the preference domain, all SCFs are  implementable whether they are BIC or not. By \autoref{WSC_BIC_functions} this entails that, in general, NBE is not a WSC solution concept.

\section{Applications} \label{applications}
The results in previous sections allow us to extend the results stemming from the  necessity of BIC for implementation to all WSC solution concepts. We take as examples three classical results from the mechanism design literature:  the impossibility of efficient bilateral trade \citep{myerson.satterthwaite.1983}, the impossibility of full surplus extraction in auctions, and the Revenue Equivalence Theorem \citep{myerson.1981}. Our results confirm that the economic intuition behind these results extends to a wide range of boundedly rational setups.

\subsection{Myerson-Satterthwaite's Impossibility Theorem} \label{MS}
\cite{myerson.satterthwaite.1983} show that efficient bilateral trade is impossible in the presence of private information, unless the planner steps in to cover some of the losses the agents face. As this result relies on the necessity of BIC for implementation in BNE, \autoref{WSC_BIC_functions} allows us to extend it to all WSC solution concepts.

As in \cite{myerson.satterthwaite.1983}, we consider a bargaining problem in which two agents (a buyer $B$ and a seller $S$) bargain over the sale of an indivisible object that each agent values at $t_i$, where each $t_i$ is distributed according to $p_i: T_i \to [0,1]$. We assume $p_i$ admits a continuous and positive pdf over the interval $[a_i, b_i]$, with $(a_S, b_S) \cap (a_B, b_B) \ne \emptyset$. We also assume  that $t_B$ is independent of $t_S$ and that each agent knows her valuation and how the valuation of the other agent is distributed. The set of alternatives consists of all pairs $(q, x)$, where $q \in [0,1]$ represents the probability that trade will happen and $x$ indicates the amount transferred from the buyer to the seller. Bernoulli utilities $u_i: A \times T_i \to \mathbb{R}$ are additively separable in money and the value of the object, and agents are risk neutral. 

Under these assumptions, \cite{myerson.satterthwaite.1983} prove that an implementing mechanism that assigns an object to the agent who values it the most is unable to ensure voluntary participation by both agents. Formally,  an SCF is \textit{ex-post efficient} if it allocates the object with probability 1 to the agent who values it the most---that is, $q(t) = 1$ whenever $t_B>t_S$, and $q(t)=0$ whenever $t_B<t_S$. Moreover, we say $f$ is \textit{individually rational} whenever $u_i(f(t),t) = q(t)t_i - x(t) \geq 0$ for all $i \in I$ and $t \in T$.

\citeauthor{myerson.satterthwaite.1983}'s \citeyearpar{myerson.satterthwaite.1983} proof 
relies on showing there exists no SCF $f$ that is simultaneously individually rational, ex post efficient, and BIC. The following corollary then follows from \autoref{WSC_BIC_functions}:

\begin{corollary}
If $f$ is individually rational and ex-post efficient, it is not fully implementable in any WSC $\sol$.
\end{corollary}

\cite{myerson.satterthwaite.1983} highlight that it is impossible to find an ex-post efficient and individually rational SCF that is  also incentive compatible for all types and agents \textit{at the same time}. This finding extends the negative results \cite{declippel.saran.serrano.2019} and \cite{crawford.2021} obtain for full implementation of SCFs in level-\emph{k} reasoning. 

\cite{kneeland.2022} shows instead it is possible to fully implement an efficient and individually rational social choice \textit{set}. Each agent can believe a different solution of the mechanism will obtain when $F$ is not a singleton, allowing the planner to decouple the incentives she provides. That is, $F$ must contain one SCF that is incentive compatible for each agent and type, but needs not contain an SCF that is incentive compatible for all types of all agents at the same time.

\subsection{Impossibility of Full Surplus Extraction} \label{surplus}
If $\sol$ is WSC,
the planner cannot implement an auction extracting all expected surplus from agents unless she excludes lower-ranked types from winning the object.

Suppose the planner is tasked with designing a mechanism to allocate a single unit of an indivisible object in exchange for the payment of a fee. Let the set of alternatives be defined as follows:
$$A = \{ (q,x) \in [0,1]^I \times \mathbb{R}^I: \sum_{i \in I} q_i \leq 1\} $$ 
That is, $f(t)$ assigns to each agent some probability of winning the object and a (non-contingent) monetary transfer. For a given $f$, denote as $q_i^f(t)$ the probability that agent $i$ receives the object and denote as $x_i^f(t)$ the associated transfer to the planner from the agent getting the object. Assume, moreover, that $T \subseteq \mathbb{R}^I$  and types are determined by a commonly known joint distribution $p: T \to (0,1)$. The value of the object to agent $i$ is determined according to a function $v_i$ that is strictly increasing in $i$'s type, and Bernoulli utilities take the additively separable form $u_i(t)= v_i(t) -x_i$.

We then say a SCS $F$ is \textit{fully extractive} whenever $x_i^f(t)= q_i^f(t)v_i(t)$ for all $t \in T$ and $f \in F$. Moreover, we say $F$ is \textit{inclusive} whenever, for all $f \in F$, $i \in I$, there exists $t_i' \in T_i$ and $t \in T$ such that $t_i' > t_i$ and $q^f_i(t)>0$. In other words, inclusivity requires that $f$ does not prevent all types $t_i$ that are ranked lower than $t_i'$ from getting the object with positive probability for all type profiles $t_{-i}$ of other agents. This is the case, for example, for ex-post efficient allocation rules.

We can then prove there exists a tradeoff between inclusivity and total surplus extraction.
\begin{corollary} \label{auctions}
If $F$ is fully extractive and inclusive, then it is not implementable in any WSC $\sol$.
\end{corollary}
The result follows because inclusivity and complete extraction of surplus entail each type has an incentive to pretend the object is worth less to her than it actually is. This creates a tension with implementability in a WSC solution concept, which implies instead there exists at least one SCF in $F$ providing each agent with the incentive not to misrepresent her type. This should be contrasted with the result in the previous example, which follows instead from the fact that the same SCF has to be simultaneously incentive compatible for all types of all agents as in the application above. The impossibility faced in this application is therefore harder to escape than the in the application in \autoref{MS}.
 
\subsection{A Revenue Equivalence Theorem}
Our results also allow us to extend 
\citeauthor{myerson.1981}'s (\citeyear{myerson.1981}) fundamental result about revenue equivalence of different auction formats to all SCFs that are fully implementable in a WSC solution concept.

As in \cite{myerson.1981}, let us assume that agents' values are drawn from set $[a_i, b_i] \subseteq \mathbb{R}^+_0$ according to some commonly known cdf $p$, that agents are risk neutral, that their utility is additively separable in money and the value of the object, and that $v_i: T_i \to \mathbb{R}^+$ is non-negative, increasing, and differentiable in $t_i$ for all $i \in I$ (this is the case, for example, if $v_i(t_i)=t_i$).

Let $\Bar{q}_i^f(t_i)$ and $\Bar{x}_i^f(t_i)$ denote respectively the average probability of winning and the  transfer for an agent of type $t_i$. An SCF $f=(q,x)$ is \textit{differentiable} if both $\bar{q}_i$ and $\bar{x_i}$ are differentiable in $t_i$ for all $i \in I$ almost everywhere, and  two SCFs $f$ and $\tilde{f}$ are \textit{assignment-equivalent} if $q^f=q^{\tilde{f}}$ almost everywhere. Notice that $f$ satisfies both conditions whenever it is ex-post efficient and   agents' values are independently and identically distributed according to a cdf $p$,  as in that case $\bar{q}(t_i) = p^{n-1}(t_i)$. 
\begin{corollary}\label{RET}
If differentiable and assignment-equivalent SCFs $f$ and $\tilde{f}$ are fully implementable in WSC $\sol$, then $\bar{x}_i^f(t_i)-\bar{x}_i^f(a_i) = \bar{x}_i^{\tilde{f}}(t_i)-\bar{x}_i^{\tilde{f}}(a_i)$ for all $i \in I$. 
\end{corollary}

\autoref{RET} establishes a generalized version of the standard Revenue Equivalence Theorem of \cite{myerson.1981}, stating that the revenue of a given SCF $f$ is determined by its allocation probability $q$ up to an additive constant $\bar{x}^f(a_i)$. If we standardize the average payment of type $a_i$ to 0, we obtain the familiar result that any two rules $f$ and $\tilde{f}$ that are fully implementable in SC $\sol$ (and their associated implementing mechanisms---for example, auctions) will yield the same ex-ante revenue to the planner unless they differ in the probability with which each type gets allocated the object. This fact entails, for example, that all ex-post efficient SCFs must yield the same revenue to the planner when $p$ is atomless.
\bibliography{support_files/references.bib}
\bibliographystyle{chicago}

\newpage
\begin{appendices}
 \section{Proofs} \label{proofs}
 \begin{proof} [Proof of \autoref{WSC_BIC_functions}]
 Suppose $f$ is implementable in $\sol$ via mechanism $\gamma=(\mu, S)$, and suppose $\sol$ is WSC for $\gamma$. Then $\sol(\gamma) \ne \emptyset$ and there exists $\sigma \in \sol(\gamma)$ such that:
 $$
 \int_{T_{-i}} u_i (\mu(\sigma(t), t) \de p_i(t_{-i}|t_i) \geq  
 \int_{T_{-i}} u_i (\mu(\sigma(t_i',t_{-i}), t) \de p_i(t_{-i}|t_i).  
 $$
 As $\sigma \in \sol(\gamma)$, implementability of $f$ yields $\mu (\sigma) = f$. Therefore, for $i \in I$ and $t_i, t_i' \in T_i$:
 $$
 \int_{T_{-i}} u_i(f(t), t) \de p_i(t_{-i}|t_i) \geq 
 \int_{T_{-i}} u_i(f(t_i',t_{-i}), t) \de p_i(t_{-i}|t_i).
 $$
 As our choice of $i, t_i$, and $t_i'$ was arbitrary, this is enough to establish that $f$ is BIC.

 Conversely, suppose  $f$ is BIC and implementable in $\sol$ via mechanism $\gamma=(\mu, S)$. Then, for all $t_i' \in T_i$ and $i \in I$: 
 $$ 
 \int_{T_{-i}} u_i (f(t), t) \de p_i(t_{-i}|t_i) \geq \int_{T_{-i}} u_i (f(t_i', t_i), t) \de p_i(t_{-i}|t_i).
 $$
 By implementability, there exists $\sigma \in \sol(\gamma)$ such that $\mu (\sigma) = f$ and thus:
 $$
 \int_{T_{-i}} u_i (\mu(\sigma(t), t) \de p_i(t_{-i}|t_i) \geq  
 \int_{T_{-i}} u_i (\mu(\sigma(t_i',t_{-i}), t) \de p_i(t_{-i}|t_i).  
 $$
 This concludes the proof.
 \end{proof}

 \begin{proof}[Proof of \autoref{BIC_SIRBIC}]
 To prove we can strengthen the result of \autoref{WSC_BIC_functions}  to SIRBIC,  we proceed by contradiction and suppose that, indeed, the incentive constraint in the proof of \autoref{WSC_BIC_functions} holds with equality. Define $\tau: T \to \Sigma$ as agreeing with $\sigma$ except that  $\tau(t)=\sigma(t_i', t_{-i})$ for all $t_{-i} \in T_{-i}$. As $\sigma$ is a solution to the mechanism, there exist $e$ such that $\sigma \in B(e)$. As $\tau$ yields the same expected utility as $\sigma$ conditional on expectations $e$, $\sigma \in B(e)$ implies $\tau \in B(e)$. Then by the definition of implementation above, for all $t_{-i} \in T_{-i}$:
 $$
 f(t_i, t_{-i})=\mu(\tau(t))=\mu(\sigma(t_i',t_{-i}))=f(t_i', t_{-i}).
 $$ 
 This concludes the proof. \end{proof}

 \begin{proof}[Proof of \autoref{WSC_BIC_sets}]
 Suppose $F$ is implementable in WSC $\sol$ via mechanism $\gamma=(\mu, S)$ with $\sol(\gamma) \ne \emptyset$. Then for each $i \in I$ and $t_i \in T_i$, there exists $\sigma \in \sol(\gamma)$ such that for all $t_i' \in T_i$:
 $$
 \int_{T_{-i}} u_i (\mu(\sigma(t), t) \de p_i(t_{-i}|t_i) \geq  
 \int_{T_{-i}} u_i (\mu(\sigma(t_i',t_{-i}), t) \de p_i(t_{-i}|t_i).  
 $$
 As $\sigma \in \sol(\gamma)$, implementability of $F$ yields $\mu (\sigma) = f$ for some $f \in F$. Therefore, for all $t_i' \in T_i$:
 $$
 \int_{T_{-i}} u_i(f(t), t) \de p_i(t_{-i}|t_i) \geq 
 \int_{T_{-i}} u_i(f(t_i',t_{-i}), t) \de p_i(t_{-i}|t_i).
 $$
 This  is enough to prove $f$ is BIC for agent $i$ and type $t_i$.

 As for the converse, suppose $F$ is implementable in $\sol$ via mechanism $\gamma$ and suppose that for all $i \in I$ and $t_i \in T_i$ there exists an $f \in F$ that is BIC for agent $i$ and type $t_i$. Then for each such $f$, $i$, and $t_i$ there exists a solution $\sigma$ such that $f = \mu (\sigma)$. A simple substitution in the BIC inequality then yields WSC holds.
 \end{proof}
 \begin{proof}[Proof of \autoref{TWSC_BIC}]
 If $F$ is implementable in $\sol$, then any $f \in F$ is such that $f = \mu (\sigma)$ for
 $\sigma \in \sol$. As $\sol$ is TWSC, it is immediate that $f$ is BIC from the definition of TWSC by substituting $f=\mu(\sigma)$. Conversely, suppose $F$ is implementable in $\sol$. As any $\sigma \in \sol$ is such that $\mu (\sigma) \in F$, TWSC follows immediately from the fact that all functions $ f \in F$ are BIC.
 \end{proof}
 \begin{proof}[Proof of \autoref{constant_expectations}]
 Suppose $F$ is implementable in SC $\sol$, and suppose $E$ is type-independent. By SC, for all $i \in I$ there exists $e \in E(\gamma)$ and $\sigma \in R(e)$ such that $(\sigma_i, e_i) \in \sol(\gamma)$. Then for all $t_i' \in T_i$,  by $\sigma \in R(e) \subseteq B(e)$ it is true that:
 $$
 \int_{T_{-i}} u_i (\mu(\sigma_i(t_i), e_i(t_{-i})), t) \de p_i(t_{-i}|t_i) \geq  
 \int_{T_{-i}} u_i (\mu(\sigma_i(t_i'), e_i(t_{-i})), t) \de p_i(t_{-i}|t_i).
 $$
 For each $i \in I$, let $f = \mu \comp (\sigma_i, e_i)$. By SC, $\mu \comp (\sigma_i, e_i) \in F$, so $f \in F$. Moreover:
 $$
 \int_{T_{-i}} u_i (f(t), t) \de p_i(t_{-i}|t_i) \geq  
 \int_{T_{-i}} u_i (f(t_i',t_{-i}), t) \de p_i(t_{-i}|t_i).
 $$
 This entails that $f$ is BIC for all types of agent $i$. This concludes the proof.
 \end{proof}

 \begin{proof}[Proof of \autoref{NBE_example}]
 Let us first derive the set of BNEs of the Maskin mechanism $(\mu, S)$ associated with $f$ under the perceived distribution of types. Let $S=\times_{i \in I} S_i$, with $S_i = (T, A, \mathbb{N})$. The outcome function $\mu$ is as follows:
 \begin{itemize} 
     \item \textbf{Rule 1:} If $s_i=(t, f(t), 0)$ for all $i \in I$, then $\mu(s)=f(t)$.
     \item \textbf{Rule 2:} If $s_j=(t', a, \mathbb{N})$ and $s_i=(t, f(t), 0)$ for all $i \ne j$, then $\mu(s)=a$ if $u_i(f(t), t)\geq a$ and $\mu(s)=f(t)$ otherwise.
     \item \textbf{Rule 3:} $\mu(s)=a$ otherwise, where $a$ is the outcome reported by the agent with the lowest index among those that reported the highest integer. 
 \end{itemize}
 As $(A, T, I)$ satisfies the economic-environment assumption, the same argument as in \cite{maskin.1999} implies that $\sigma(t)$ does not fall under Rule 2 or Rule 3 for all $t \in T$. If it were otherwise, at least one agent could report a higher integer than her opponents and achieve $a \in A$ such that $u_i(a,t)>u_i(\mu(\sigma(t)),t)$. Therefore, for all states $t \in T$, $\sigma(t)$ falls under Rule 1. 

As $\sigma(t)$ falls under Rule 1 for all $t\in T$,  there exists a $t' \in T$ such that $\sigma_i(t_i)=(t', f(t'), 0)$ for all $t_i \in T_i$ and $i \in I$. For the sake of contradiction, suppose now that $t' \ne t$. As $\sigma$ is an equilibrium, it must hold for all $i \in I$ that $u_i(f(t'), t) \geq u_i(a, t)$ for all $a \in A$ such that $u_i(f(t'), t') \geq u_i(a, t')$, as otherwise at least one agent $i$ could play $s_i' = (t', a, 0)$ and obtain $u_i(a, t) > u_i(f(t'), t)$. Notice now that, by single crossing, for all $i \in I$ there exists a $z \in A$ such that $u_i(f(t'),t') \geq u_i(z, t')$ and $u_i(z, t)> u_i(f(t'), t)$. This entails it would be profitable for type $i$ to deviate to $s_i'=(t, z, 0)$, contradicting the premise that $\sigma$ is an equilibrium strategy. Therefore, it must be $\sigma(t)= (t, f(t), 0)$ for all $i \in I$ and $t \in T$.
 
 It remains to be shown that no type $t_i$ of each $i \in I$ has a profitable deviation from $\sigma(t)$. As $i$ can induce $a$ only if $u_i(\mu(\sigma(t)), t) \geq u_i(a, t)$, this concludes the proof.
 \end{proof}

 \begin{proof}[Proof of \autoref{auctions}]
 We now show that supposing $F$ is implementable in WSC $\sol$ leads to a contradiction. Consider any agent $i \in I$. By inclusivity, there exist types $t_i, t_i' \in T_i$ such that $q^f_i(t_i', t_{-i})>0$ and $t_i>t_i'$. By WSC and \autoref{WSC_BIC_sets}, we then know that if $F$ is implementable in $\sol$, for all $i \in I$ and $t_i \in T_i$, then there exists an $f \in F$ that is BIC for $i$ and $t_i$. Therefore, for all $i \in I$, $t_i \in T_i$ and $t_i'<t_i$, full surplus extraction implies:
 $$
 0= \int_{T_{-i}} (q_i^f(t)v_i(t) - q_i^f(t)v_i(t)) \de p_i(t_{-i}| t_i) \geq \int_{T_{-i}} q_i^f(t_i',t_{-i})(v_i(t) - v_i(t_i',t_{-i})) \de p_i(t_{-i}| t_i)
 $$
 As $v_i$ is strictly increasing in $i$'s type and $F$ is inclusive:
 $$
 \int_{T_{-i}} q_i^f(t_i', t_{-i}) (v_i(t)-v_i(t_i', t_{-i})) \de p_i(t_{-i}| t_i)>0. 
 $$
 This inequality contradicts the fact that $f$ is BIC for $i$ and $t_i$, concluding the proof.
 \end{proof}
 
 \begin{proof}[Proof of \autoref{RET}]
 As $f$ is implementable in $\sol$ WSC, it is BIC. So $t_i^*=t_i$ must maximize the payoff function $\bar{q}^f(t_i^*) v_i(t_i) - \bar{x}_i^f(t_i^*)$.
 A necessary condition for a maximum is that the first derivative with respect to $v_i$ of this function is null at $t_i$---that is,  $\frac{\partial \bar{x}^{f}(t_i)}{\partial t_i}
 =\frac{\partial \bar{q}^{f}(t_i)}{\partial t_i} v_i(t_i)$.
Then:
 $$
 \bar{x}(t_i) - \bar{x}(a_i) = \int_{a_i}^{t_i}  v_i(t_i') \frac{\partial \bar{q}^f(t_i')}{\partial t_i}\de t_i'.
 $$
 Analogous reasoning for $\tilde{f}$ and assignment-equivalence yield $
 \bar{x}^f(t_i)-\bar{x}^f(a_i) = \bar{x}^{\tilde{f}}(t_i)-\bar{x}^{\tilde{f}}(a_i)$, concluding the proof.
 \end{proof}
 \newpage

\section{Extensions of the Model} \label{extensions}
 \subsection*{Non-standard Choice Correspondences} \label{choice_functions}
 This section relaxes the assumption that agents best respond to their expectations, generalizing the results in the main body of the paper beyond the domain of von Neumann–Morgenstern preferences. 

 We can interpret the revelation principle as saying that some lotteries in the choice sets  induced by an indirect implementing mechanism  (but not in the direct one) can be safely neglected, as they are not going to be relevant. Formally, this requires that restricting the choice set of an agent of type $t_i$ to the set of lotteries that would be a solution to the mechanism for some type $t_i' \in T_i$ does not affect her choice. This will require us to impose some form of Contraction Consistency, or Independence of Irrelevant Alternatives (see, for example, Property $\alpha$ of \cite{sen.1971}). In the argument below, we  only maintain the assumption that agents are consequentialist---that is, that their choices depend only on the set of alternatives they choose from.\footnote{This rules out, for example, preferences for truth-telling.}

As in \cite{saran.2011} and \cite{barlo.dalkiran.2022}, we  model individual strategic decisions, for all $i \in I$, as choices over a set of \textit{interim Anscombe-Aumann acts} (IAA acts) $x_i: T_{-i} \to \Delta(A)$.  Denote as $\mathcal{X}$ the set of all IAA acts.

We can then define a choice correspondence $C_{i, t_i}$ as mapping each non-empty subset $X$ of $\mathcal{X}$ to a subset of $\Delta(X)$. That is, for all $X \subseteq \mathcal{X}$, $C_{i, t_i}(X) \subseteq \Delta(X)$. As in \cite{barlo.dalkiran.2022} and unlike in \cite{saran.2011}, we do not assume $C_{i, t_i}$ is generated by a menu-dependent preference order. 

 Notice that for any given $s_i \in S_i$ and $\sigma_{-i} \in \Sigma_{-i}$, the function $\mu(s_i, \sigma_{-i})$ is an IAA act. We can then denote the set of acts agent $i$ of type $t_i$ chooses from given her expectations as follows:
 $$
 O_i(\sigma_{-i}) = \left\{ x_i \in \mathcal{X}: x_i = \mu(s_i, \sigma_{-i}), s_i \in S_i \right\}
 $$
 As in previous sections, we say $s_i \in \Delta(S_i)$ is a reply to $\sigma_{-i}$ for type $t_i$ whenever $s_i \in R_{i, t_i}(\sigma_{-i}) \subseteq \mu^{-1}(C_{i, t_i}(O_i(\sigma_{-i})))$. That is, the outcome of the strategies chosen as a response to $\sigma_{-i}$ is a subset of 
 what agent $i$ of type $t_i$ would choose from the set of acts $O_i(\sigma_{-i})$. Notice that $\mu^{-1}(C_{i, t_i}(O_i(\sigma_{-i})))$ coincides with the set $B_{i,t_i}$ considered in the main text in the case in which the agent maximizes expected utility given $\sigma_{-i}$. We moreover say $\sigma$ is a solution to a mechanism $\gamma$ whenever there exists $e \in E(\gamma)$ such that $\sigma \in R(e)$. 

 Let $O_i^{f,t_i}$ denote the set of IAA acts that agent $i$ can generate in the direct mechanism $(f,T)$ when her opponents truthfully report their type:
 $$
 O_i^{f,t_i} = \left\{ x_i \in \mathcal{X}: x_i \in f(t_i', \id_{-i}) \text{ where } t_i' \in T_i \right\}
 $$
 Incentive Compatibility can then be generalized as in \cite{saran.2011}:

 \begin{definition}[Incentive Compatibility (IC)]
 Let $C_{i, t_i}$ be given. We say $f$ satisfies IC for type $t_i \in T_i$ and $i \in I$ whenever $f(t_i,\cdot) \in C_{i, t_i}(O_i^{f,t_i})$. We say $f$ is IC whenever it is IC for all $t_i \in T_i$ and $i \in I$.
 \end{definition}

 In other words, we require agents to choose the act associated with their type $t_i$ when they expect their opponents to choose the acts associated with their types as well. In the case of BIC, this coincides with the set of acts maximizing expected utility in the choice set. To derive our main result for this section, we redefine WSC in terms of choice correspondences rather than utility maximization.
 \begin{definition}[Weak Choice Consistency (WCC)] 
 We say a solution concept $\sol$ satisfies WCC for a class of mechanisms $\tilde{\Gamma}\subseteq \Gamma$ whenever for all $\gamma \in \tilde{\Gamma}$, $i \in I$, $t_i \in T_i$ there exists $\sigma \in \sol(\gamma)$ such that 
 $\mu(\sigma_i(t_i), \sigma_{-i}) \in C_{i, t_i} (X_i(\sigma_{-i}))$, where:
 $$
 X_i(\sigma_{-i}) = \{x_i\in \mathcal{X}: x_i=\mu(\sigma_i(t_i'), \sigma_{-i}) \text{ with } t_i' \in T_i \}.
 $$
 We say $\sol$ satisfies WCC if it satisfies WCC for all $\gamma \in \Gamma$ such that $\sol(\gamma)\ne \emptyset$.
 \end{definition}
 It is immediately possible to extend \autoref{WSC_BIC_sets} and, when $F$ is a singleton, \autoref{WSC_BIC_functions}. 
 \begin{theorem} \label{WCC_IC}
 If $F$ is implementable in WCC $\sol$, then for all $i \in I$ and $t_i \in T_i$ there exists an $f^{i, t_i} \in F$ that is IC for $i$ and $t_i$. Conversely, if $F$ is implementable and there exists an $f^{i, t_i} \in F$ that is IC for $i$ and $t_i$, then $\sol$ is WCC for $\Gamma^F$.
 \end{theorem}

 Notice that both WSC and WCC implicitly assume a mild form of contraction consistency between choices in $O_i(\sigma_{-i})$ and in $X_i(\sigma_{-i})$. In fact, if $\sigma \in \sol(\gamma)$ is such that $\mu(\sigma_i(t_i), \sigma_{-i}) \in C_{i,t_i}(O_i(\sigma_{-i}))$ and $X_i(\sigma_{-i}) \subseteq O_i(\sigma_{-i})$, WCC entails that $\mu(\sigma_i(t_i), \sigma_{-i}) \in C_{i, t_i}(X_i(\sigma_{-i}))$. 

 This implicit assumption means it is not as easy to provide a sufficient condition for WCC as it was for WSC. Let us parallel the definition of SC and say $\sol$ is \textit{Choice Consistent} (CC) for mechanism $\gamma$ whenever there exist $e, e' \in E(\gamma)$ and $\sigma \in R(e)$ such that $(\sigma_i, e_{i, t_i}) \in R(e')$. While this entails $\mu(\sigma_i(t_i), e_{i,t_i}) \in C_{i, t_i}(O_i(e_{i,t_i}))$, this is not enough to establish WCC, as it does not preclude the possibility that $\mu(\sigma_i(t_i), e_{i,t_i}) \not\in C_{i, t_i}(X_i(e_{i,t_i}))$. Without any form of contraction consistency, CC just implies that for all $i \in I$ and $t_i \in T_i$ there exists an $f \in F$ and $\mathcal{O} \subseteq \mathcal{X}$ such that $O_i^{f, t_i} \subseteq \mathcal{O}$ and $f(t_i, \cdot) \in C_{i, t_i}(\mathcal{O})$. Using \citeauthor{barlo.dalkiran.2022}'s \citeyearpar{barlo.dalkiran.2022} terminology, we can say $f$ is \textit{quasi-incentive compatible} (QIC) for agent $i$ of type $t_i$. 

 \cite{chernoff.1954} provides an example of a class of choice correspondences ruling out such a possibility without implying maximization of rational preferences \citep{sen.1971}. We say a choice correspondence $C_{i, t_i}$ satisfies Independence of Irrelevant Alternatives (IIA) whenever for all $X, Y \subseteq \mathcal{X}$:
 $$
 C_{i, t_i}(X) \subseteq Y \subseteq X \implies C_{i, t_i}(X) \subseteq C_{i, t_i}(Y).
 $$
 It is  easy to see that IIA entails that if $\sol(\gamma)$ is CC, it is WCC as $\mu(\sigma_i(t_i), e_{i, t_i}) \in C_{i, t_i}(O_i(e_{i, t_i}))$ and:
 $$
 C_{i, t_i}(O_i(e_{i, t_i})) \subseteq X_i(e_{i, t_i}) \subseteq O_i(e_{i, t_i}) \implies 
 C_{i, t_i}(O_i(e_{i, t_i}))\subseteq C_{i, t_i}(O_i^{f, t_i}).  
 $$
It is  immediately possible to derive the following by the same argument as in the main text. 
 \begin{corollary}
 If $F$ is implementable in CC $\sol$, then there exists an $f^{i, t_i} \in F$ that is QIC for type $t_i$ and agent $i$.  If, moreover, $C_{i, t_i}$ satisfies IIA, $f^{i, t_i}$ is IC for type $t_i$ and agent $i$. 
 \end{corollary}

Even if non-equilibrium models with non-rational choice correspondences have not been considered in the literature, these results allow us to extend the findings of \cite{declippel.saran.serrano.2019} and \cite{kunimoto.saran.serrano.2020} about level-$k$ and rationalizable implementation to all consequentialist choice correspondences.\footnote{\cite{barlo.dalkiran.2022} already extend BNE to non-rational choice correspondences with their Behavioral Interim Equilibrium (BIE) solution concept.} To this end, it is enough to tweak the definitions of the solution concepts in the main text by replacing the assumption $R \subseteq B$ with $R \subseteq \mu^{-1}(C_{i, t_i}(O_i(\sigma_{-i})))$. In particular, we can say IC is necessary for implementation in these solution concepts whenever $C$ is IIA.

A limitation of this analysis is in the assumption that agents' choices depend only on the menu of acts they choose from. This assumption rules out, for instance, Quantal Response Equilibrium\citep{mckelvey.palfrey.1995} and Sampling Equilibrium \citep{osborne.rubinstein.1996}). In these models, agents' choices depend not only on the menu of available acts  but on the number of times an act appears in the menu. We can accommodate these models by relaxing the assumption that, for all $i$ and $t_i$, $C_{i, t_i}$'s domain is the set of all non-empty \textit{sets} $X \subseteq \mathcal{A}$. Let us assume instead that $C_{i, t_i}$'s domain is the set of all non-empty \textit{bags} $X$ with support in $\mathcal{A}$.\footnote{A bag, or multiset, is a generalization of the concept of set that allows more than one instance of each element. Its support is the set of elements that appear at least once.} We can get the same result as above (with slightly heavier notation) by  adjusting the definitions of IC, QIC, WCC, CC, and IIA accordingly.  In this case, both Sampling Equilibrium and Quantal Response Equilibrium can be considered as special cases of Behavioral Interim Equilibrium \citep{barlo.dalkiran.2022}. 

% here I could add an hybrid between a sampling equilibrium and a level-k reasoning model as an example (each agent samples k times, and has a belief 

Models as Fairness Equilibrium \citep{rabin.1993} are ruled out by our formulation too, as agents' choices may depend on the menu of acts \textit{their opponents} are choosing from. Again, redefining the domain of the choice correspondence to be the set of all collections of opportunity sets (one for each $i \in I$) yields a special case of Behavioral Interim Equilibrium.

\subsection*{Epistemic Foundations} \label{epistemic_argument}
 To better appreciate how restrictive the conditions for the necessity of BIC are, this section provides an epistemic justification for (Weak) Solution Consistency. The bulk of the argument is based on the one \cite{dekel.fudenberg.morris.2007} propose for characterizing ICR. SC obtains whenever each type of each agent is rational, knows the type space, and knows she could successfully mimic other types. This makes the overall requirements for SC rather weak and the related class of solution concepts rather large. 

 Let $\gamma$ be fixed, $T$ be finite, and let $P_i(t_i') = \{ t \in T: p_i(t_{-i}|t_i')>0 \}$. Denote as $\sol(\gamma, t)$ the set of profiles $s$ such that there exists $\sigma \in \sol(\gamma)$ such that $\sigma(t)=s$. For the purpose of this section, we assume $\Sigma= \times_{i \in I} \Sigma_i$. While this rules out some solution concepts, it still captures most of the solution concepts discussed in the main text, and it makes the exposition in the proof significantly simpler.

 Let $H_i$ be a finite set of epistemic types for player $i$, and let $H= \times_{i \in I} H_i$.\footnote{Finiteness of $S$, $T$, and $H$ is not necessary for the argument, but it simplifies the exposition by avoiding the use of measure-theoretic notation.} An epistemic model specifies the following:
 \begin{itemize}
     \item functions $\phi_i: H_i \to \Delta(H_{-i})$, mapping $i$'s epistemic type to a belief over others' epistemic types
     \item $i$'s epistemic strategy $\eta_i:H_i \to \Delta(S_i)$, assigning a distribution over actions to each epistemic type
     \item $i$'s standard type $\tau_i: H_i \to T_i$
 \end{itemize}
 An epistemic model is then a tuple $(H_i, \phi_i, \eta_i, \tau_i)_{i \in I}$ with state space $H$. For a given $\phi_i$, denote the joint distribution over opponents' actions and standard types as follows:
 $$
 \lambda_i(h_i)[s_{-i}, t_{-i}]=  \int_{H_{-i}} \eta_{-i}(h_{-i}')[s_{-i}] \indicator_{\{\tau_{-i}(h_{-i}')=t_{-i})\}}(h_{-i}') \de \phi_i(h_i)[h_{-i}']
 $$
 For all $s_{-i} \in \Delta(S_{-i})$, let $\lambda_i(h_i, t_{-i})[s_{-i}] = \lambda_i(h_i)[s_{-i}, t_{-i}]$ denote the probability profile $s_{-i}$  is played conditional on type profile $t_{-i}$. We then define the event in which $i$'s beliefs over her opponents' standard types are consistent with $\phi_i$ as follows:
 $$
 W_i = \{ h \in H: \lambda_i(h_i)[S_{-i}, \tau_{-i}(h_{-i})] = p_i(\tau_{-i}(h_{-i})|\tau_i(h_i)) \}
 $$
 Notice that $h \in W_i$ implies $\lambda_{i}(h_i, \cdot): T_{-i} \to \Delta_{-i} (S_{-i})$, so that $\lambda_i(h_i, \cdot) \in \Sigma_{-i}$ is a properly defined expectation. Let $RAT_i$ be the set of states such that, conditional on beliefs $\phi(h_i)$, playing the mixed action associated with  one's own epistemic type $h_i$ yields higher expected utility than playing the action associated with any other epistemic type $h_i'$:
 $$
 RAT_i = \left\{h \in H: \eta_i(h_i) \in \arg\max_{h_i' \in H_i} \int_{H_{-i}} u_i(\mu(\eta(h_i',h_{-i}))) \de \phi_i(h_i)[h_{-i}] \right\}
 $$
 We now define two more events. The first event can be interpreted as the set of epistemic states such that $i$ playing the action associated with her epistemic type is a solution to the mechanism if her opponents do the same:
 $$
 TT_i= \{h \in H: (\eta_{i}(h_i), \lambda_i(h_i, \tau_{-i}(h_{-i}))) \cap \sol(\gamma, \tau(h)) \ne \emptyset  \}
 $$
 The second event can be interpreted as the set of epistemic states $h$ such that $i$ playing the action associated with an epistemic type different from her own is a solution to the mechanism when her opponents play the action associated with their epistemic type:
 $$
 PM_i= \{h \in H: (\eta_{i}(\tau_i^{-1}(t_i')), \lambda_i(h_i, \tau_{-i}(h_{-i}))) \cap \sol(\gamma, (t_i',\tau_{-i}(h_{-i})) \ne \emptyset, t_i' \ne \tau_i(h_i)  \}
 $$
 This drives the necessity of (partial) BIC for full implementation: if type $t_i$ can get away with mimicking any other type $t_i'$, it   becomes necessary to provide her with incentives not to do so by choosing a BIC SCF $f$. Formally, for any $t_i' \in T_i$, $\eta_{i}(\tau_i^{-1}(t_i'))$ represents the set of all actions played by epistemic types with standard type $t_i'$. So  $(\eta_{i}(\tau_i^{-1}(t_i')), \lambda_i(h_i, \tau_{-i}(h_{-i}))$ corresponds to all distributions over actions that $i$ can induce by mimicking some type $t_i' \ne \tau_i(h_i)$, conditional on her expectations about the actions of her opponents. The requirement that at least one profile in $(\eta_{i}(\tau_i^{-1}(t_i')), \lambda_i(h_i, \tau_{-i}(h_{-i}))$ must belong to  $\sol(t_i',\tau_{-i}(h_{-i})$ can  be interpreted by considering that, in order to successfully mimic a different type, $t_i$ must also make the planner  think that the profile of actions the planner observes comes from type profile $(t_i', \tau_{-i}(h_{-i}))$ rather than $(t_i, \tau_{-i}(h_{-i}))$.  

 Finally, we denote the intersection of the last two events as follows:
 $$
 SOL_i = TT_i \cap PM_i
 $$
 This leads to a second possible interpretation of the two events above: as in \cite{zambrano.2008}, we can take them as meaning that the action $\eta_i$ prescribes for $i$, together with her expectations, will form a solution to the game. \cite{zambrano.2008} uses mutual knowledge of this event to characterize the set of correlated rationalizable action profiles. Notice that the characterization in \cite{zambrano.2008}, unlike others, relies only on \textit{mutual} rather than \textit{common} knowledge. 

 We can now prove that WSC is almost characterized by the existence of an epistemic type for each type $t_i \in T_i$ that is rational, that  knows the standard type space and that  that mimicking is possible.\footnote{The property discussed in this section is weaker than WSC, as the fact that for all $t \in P_i(t_i)$ there exists $\tilde{\sigma}(t)$ such that $\Tilde{\sigma}(t)=\sigma(t)$ does not generally imply $\sigma^{i,t_i} \in \sol(\gamma)$. Such a gap is immaterial for the results of \autoref{incentive_compatibility}, but it would require the use of heavier notation.}
 \begin{theorem} \label{epistemic_wsc}
 The two following statements are equivalent:
 \begin{enumerate}
     \item There exists an epistemic model such that for all $i \in I$ and $t_i \in T_i$ there exists $h^* \in RAT_i \cap K_i(W_i \cap SOL_i)$ with $\tau_i(h_i^*)=t_i$.
     \item For all $i \in I$ and $t_i \in T_i$ there exists $\sigma^{i, t_i} \in \Sigma$ such that  $\sigma^{i, t_i}(\tilde{t}) \in \sol(\gamma,t)$ for all $\tilde{t} \in P_i(t_i)$ and for all $t_i'\in T_i$:
 $$
 \int_{T_{-i}} u_i(\mu(\sigma^{i, t_i}(t))) \de p_i(t_{-i}|t_i) \geq \int_{T_{-i}} u_i(\mu(\sigma^{i, t_i}(t_i',t_{-i}))) \de p_i(t_{-i}|t_i)
 $$.
 \end{enumerate}
 \end{theorem}
 We can apply a similar argument to SC  to appreciate how different it is from WSC. The main difference arises because utility maximization does not characterize the set of responses anymore  and because such a set may depend on the full profile of expectations $e$. For any profile $\tilde{\sigma}_{-i} \in \Sigma_{-i}$, let $R_{i, t_i}(\tilde{\sigma}_{-i})$ denote the set of profiles $\sigma$ that are a response to an expectation profile $e$ consistent with type $t_i$ of agent $i$ who expects her opponents to play $\tilde{\sigma}_{-i}$. Formally:
 $$
 R_{i, t_i}(\tilde{\sigma}_{-i}) = \{ s_i \in \Delta(S_i): s_i = \sigma(t) \text{ for } \sigma \in R(e) \text{ s.t. } e_{i, t_i} = \tilde{\sigma}_{-i} \}.
 $$
 We can then define:
 $$
 RAT_i^* = \{h \in H: \eta_i(h_i) \in R_{i,\tau_i(h_i)}(\lambda_i(h_i, \cdot))\}.
 $$
 We can finally characterize SC as follows.
 \begin{theorem} \label{epistemic_rc}
 The  following two statements are equivalent:
 \begin{enumerate}
     \item There exists an epistemic model such that for all $i \in I$ and $t_i \in T_i$ there exists $h^* \in RAT_i^* \cap K_i(W_i \cap SOL_i)$ with $\tau_i(h_i^*)=t_i$.
     \item For all $i \in I$ and $t_i \in T_i$ there exists $e \in E(\gamma)$ and $\sigma \in \Sigma$ such that $\sigma_i(t_i) \in R_{i,t_i}(e_{t_i})$ and $(\sigma_i, e_{i, t_i})(\tilde{t}) \in \sol(\gamma, \tilde{t})$ for all $\tilde{t} \in P_i(t_i)$.
 \end{enumerate}
 \end{theorem}

 The main difference between the two conditions does not lie in what agents know but rather in the assumptions made about the way they respond to their expectations. This finding highlights the central role played by knowledge of $SOL_i$ in determining whether a solution concept is WSC or SC. Cursed Equilibrium in \autoref{wsc_examples} is a case in point. In that example, even if in all solutions each type of each agent plays a pure strategy, $\Bar{\sigma}_{-i}$ assigns equal probability to both actions. Thus $(\sigma_i(t_i), \Bar{\sigma}_{-i}(t_{-i}))$  fails to be a solution of the mechanism for any type profile $t$. 

 \subsection*{Proofs for \autoref{extensions}}

 \begin{proof}[Proof of \autoref{WCC_IC}]
 Suppose $F$ is implementable in WCC $\sol$ via mechanism $\gamma=(\mu, S)$ with $\sol(\gamma) \ne \emptyset$. Then for each $i \in I$ and $t_i \in T_i$ there exists $\sigma \in \sol(\gamma)$  such that 
 $\mu(\sigma_i(t_i), \sigma_{-i}) \in C_{i, t_i} (X_i(\sigma_{-i}))$, where:
 $$
 X_i(\sigma_{-i}) = \{x_i \in \mathcal{X}: x_i=\mu(\sigma_i(t_i'), \sigma_{-i}) \text{ with } t_i' \in T_i \}.
 $$
 As $\sigma \in \sol(\gamma)$, implementability of $F$ yields $\mu (\sigma) = f$ for some $f \in F$. Then:

 $$
 X_i(\sigma_{-i}) = \{x_i \in \mathcal{X} : x_i=\mu(\sigma_i(t_i'), \sigma_{-i})=f(t_i', \cdot), t_i' \in T_i \} = O_i^{f, t_i}.
 $$
 Then $f(t_i, \cdot) = \mu(\sigma_i(t_i), \sigma_{-i}) \in C_{i, t_i} (X_i(\sigma_{-i})) = C_{i, t_i} (O_i^{f, t_i})$, which is enough to prove $f \in F$ is BIC for agent $i$ and type $t_i$.

 As for the converse, suppose $F$ is implementable in $\sol$ via mechanism $\gamma$ and suppose that for all $i \in I$ and $t_i \in T_i$ there exists an $f \in F$ that is IC for agent $i$ and type $t_i$. As $f$ is IC for $i$ and $t_i$, $f(t_i, \cdot) \in O_i^{f, t_i}$. Moreover, as $F$ is implementable, there exists a solution $\sigma \in \sol(\gamma)$ such that $f = \mu (\sigma)$. Then:
 $$
 O_i^{f,t_i}= \{x_i \in \mathcal{X}: x_i=f(t_i', \cdot)=\mu(\sigma_i(t_i'), \sigma_{-i}) \text{ with } t_i' \in T_i \} = X_i(\sigma_{-i}).
 $$
 As our initial choice of $i$ and $t_i$ was arbitrary, this concludes the proof.
 \end{proof}

 \begin{proof}[Proof of \autoref{epistemic_wsc}]
 $1 \implies 2:$ Let $h^* \in  RAT_i \cap K_i(W_i \cap SOL_i)$. We invoke the following Lemma, which is proven separately.

 \begin{lemma}\label{first_lemma}
     If $h^* \in K_i(W_i \cap SOL_i)$, there exists $z_i: T \to H_i$ such that $\tau_i(z_i(t))=t_i$ for all $t_i \in T_i$ and $(\eta_i(z_i(t)), \lambda_i(h_i^*, t_{-i})) \in \sol(\gamma, t)$ for all $t \in P_i(\tau_i(h_i^*))$.
 \end{lemma}
 Set  $\sigma(t) = (\eta_i(z_i(t)), \lambda_i(h_i^*, t_{-i}))$ for all $t \in T$. Then $h^* \in RAT_i$ yields that for all $h_i' \in H_i$:
 $$
 \int_{H_{-i}} u_i(\mu(\eta(h_i^*, h_{-i})) \de \phi_i(h_i^*)[h_{-i}] \geq
 \int_{H_{-i}} u_i(\mu(\eta(h_i', h_{-i}))) \de \phi_i(h_i^*)[h_{-i}].
 $$
 This entails that for all $t_i' \in T_i$:
 $$
 \int_{H_{-i}} u_i(\mu(\eta(h_i^*, h_{-i})) \de \phi_i(h_i^*)[h_{-i}] \geq
 \int_{H_{-i}} u_i(\mu(\eta(z_i(t_i', t_{-i}), h_{-i}))) \de \phi_i(h_i^*)[h_{-i}].
 $$
 By Fubini-Tonelli and our choice of $\sigma$ we can rewrite the inequality above as:
 $$
 \int_{T_{-i}} u_i(\mu(\sigma(\tau_i(h_i^*), t_{-i})), t) \de p_i(t_{-i}|t_i) \geq \int_{T_{-i}} u_i(\mu(\sigma(t_i',t_{-i})), t) \de p_i(t_{-i}|t_i).
 $$
 This concludes the proof.

 $2 \implies 1$: For all $i \in I$ and $t_i \in T_i$, let $H_j$ consist of all pairs $(s_j, t_j)$ such that $s_j= \sigma_j^{i, t_i}$.\footnote{As $T$ is finite,  $H$ is finite as well.} Let $\eta_j(h_j)=\eta_j(s_j, t_j)=s_j$ and $\tau_j=\tau_j(s_j, t_j)=t_j$. For all $j \in I$ and $h_j \in H_j$, let:
 $$
 \phi_j(h_j)[h_{-j}] = \phi_j(s_j,t_j)[(s_{-j},t_{-j})] = \sigma_{-j}^{j, t_j}(t_{-j})[s_{-j}] p_j(t_{-j}|t_j).
 $$
 We now show that this epistemic model is such that for all $i \in I$ and $t_i \in T_i$ there exists $h^* \in RAT_i \cap K_i(W_i \cap SOL_i)$ with $\tau_i(h_i^*)=t_i$. 

 Fix now any $i \in I$ and $t_i \in T_i$, and consider any $h^*$ such that $h_i^*=(\sigma_i^{i, t_i}(t_i), t_i)$. By our choice of $\tau$, we can see that $\tau_i(h_i^*)=t_i$.

 We first show $h^* \in K_i(W_i)$. As $\sigma_{-i}^{i, t_i}(t_{-i})[S_{-i}]=1$, for all $t_{-i} \in T_{-i}$ we have the following:
 $$
 \int_{H_{-i}} \indicator_{\{\tau_{-i}(h_{-i})=t_{-i}\}}(h_{-i})
 \de\phi_i(h_i)[h_{-i}] =\phi_i(s_i,t_i)[(S_{-i},t_{-i})] = p_i(t_{-i}|t_i)
 $$
 Moreover, $h^* \in RAT_i$. Notice first that for all $h_i' \in H_i$, there exists a $t_i' \in T_i$ such that $\eta_i(h_i')=\sigma_i^{i,t_i}(t_i')$  by our definition of $H_i$. By assumption:
 $$
 \int_{T_{-i}} u_i(\mu(\sigma^{i, t_i}(t)), t) \de p_i(t_{-i}|t_i) \geq \int_{T_{-i}} u_i(\mu(\sigma^{i, t_i}(t_i', t_{-i}), t) \de p_i(t_{-i}|t_i).
 $$
 We can use Fubini-Tonelli to rewrite the inequality above as follows:
 $$
 \int_{H_{-i}} u_i(\mu(\eta(h_i^*, h_{-i})) \de \phi_i(h_i^*)[h_{-i}] \geq
 \int_{H_{-i}} u_i(\mu(\eta(h_i', h_{-i})) \de \phi_i(h_i^*)[h_{-i}]
 $$
 This is enough to prove $h^* \in RAT_i$. 

  Suppose now for the sake of contradiction that $\phi_i(h_i^*)[SOL_i]<1$, so that $h^* \not \in K_i(SOL_i)$. Then there exists an epistemic state $h_{-i}$ with $\phi_i(h_i^*)[h_{-i}]>0$ such that either:
 $$
 (\eta_{i}(h_i^*), \lambda_i(h_i^*, \tau_{-i}(h_{-i})) \not \in \sol(\gamma, \tau(h_i^*, h_{-i})) 
 $$
 or, for some standard type $t_i'$ and we have that for all $h_i$ with $\tau_i(h_i)$:
 $$
 (\eta_{i}(h_i), \lambda_i(h_i^*, \tau_{-i}(h_{-i})) \not \in \sol(\gamma, (t_i',\tau_{-i}(h_{-i})).
 $$
 In either case, $\phi_i(h_i^*)[h_{-i}]>0$ implies $p_i(\tau_{-i}(h_{-i})| t_i)>0$ and thus either: 
 $$(\sigma_i^{i, t_i}(t_i), \lambda_i(h_i^*, \tau_{-i}(h_{-i})) = \sigma^{i, t_i}(t_i, \tau_{-i}(h_{-i})) \not\in \sol(\gamma, (t_i,\tau_{-i}(h_{-i}))$$ 
 or
 $$(\sigma_i^{i, t_i}(t_i'), \lambda_i(h_i^*, \tau_{-i}(h_{-i})) = \sigma^{i, t_i}(t_i', \tau_{-i}(h_{-i})) \not\in \sol(\gamma, (t_i',\tau_{-i}(h_{-i})).$$
 This contradicts our premises, concluding the proof.
 \end{proof}

 \begin{proof}[Proof of \autoref{epistemic_rc}]
 $1 \implies 2:$ Let $h^* \in  RAT_i^* \cap K_i(W_i \cap SOL_i)$. We invoke again \autoref{first_lemma} to argue that there exists a $z_i: T \to H_i$ such that $\tau_i(z_i(t))=t_i$ for all $t_i \in T_i$ and $(\eta_i(z_i(t)), \lambda_i(h_i^*, t_{-i})) \in \sol(\gamma, t)$ for all $t \in P_i(\tau_i(h_i^*))$.

 Consider now profile $(\eta_i(z_i(t)), \lambda_i(h_i^*, t_{-i}))$ for all $t_i \in T_i$. From $h^* \in RAT_i^*$  there exists an $e \in E(\gamma)$ and a $\sigma \in R(e)$ such that $e_{i,t_i}=\lambda_i(h_i, \cdot)$ and $\eta_i(h_i^*)=\sigma_i(\tau_i(h_i^*))$.  We can then rewrite these equations as follows:
 $$
 (\sigma_i, e_{i, t_i})(t) =
 (\eta_i(z_i(t)), \lambda_i(h_i^*, t_{-i}))
 $$
 By construction of $z_i$, for all $t \in P_i(\tau_i(h_i^*))$:
 $$
 (\sigma_i, e_{i, t_i})(t) =
 (\eta_i(z_i(t)), \lambda_i(h_i^*, t_{-i})) \in \sol(\gamma, t).
 $$
 This concludes the proof.

 $2 \implies 1$: The proof is analogous to the proof of \autoref{epistemic_wsc}, except  we now have to show $h^* \in RAT_i^*$. This follows easily, as $\sigma \in R(e)$, $e_{i,t_i}=\lambda_i(h_i^*, \cdot)$ and $\eta_i(h_i^*)=\sigma_i(t_i)$. 
 \end{proof}

 \begin{proof}[Proof of \autoref{first_lemma}]
 We invoke the following Lemma, proved below:
 \begin{lemma}\label{second_lemma}
 If $h^* \in K_i(W_i \cap SOL_i)$, for all $t_{-i} \in T_{-i}$ such that $p_i(t_{-i}|\tau_i(h_i^*))>0$
 there exists an $h_{-i} \in H_{-i}$ such that $\tau_{-i}(h_{-i})=t_{-i}$ and $(h_i^*, h_{-i}) \in SOL_i$.
 \end{lemma}

 Let $t \in T$ be such that $p_i(t_{-i}|\tau_i(h_i^*))>0$ (that is, $t \in P_i(\tau_i(h_i^*))$), and let $(h_i^*, h_{-i}) \in SOL_i$ be such that $\tau_{-i}(h_{-i})=t_{-i}$ (we know such an $h_{-i}$ exists by \autoref{second_lemma}). 
Recall that:
 $$
 SOL_i= \{h \in H: (\eta_{i}(\tau_i^{-1}(t_i)), \lambda_i(h_i, \tau_{-i}(h_{-i}))) \cap \sol(\gamma, (t_i,\tau_{-i}(h_{-i})) \ne \emptyset, t_i \in T_i  \}.
 $$
 As $(h_i^*, h_{-i}) \in SOL_i$, there exists then an $h_i\in \tau_i^{-1}(t_i)$ such that:
 $$
 (\eta_{i}(h_i), \lambda_i(h_i^*, t_{-i})) \cap \sol(\gamma, t) \ne \emptyset.
 $$
 For all $t \in P_i(\tau_i(h_i^*))$ with $t_i\ne \tau_i(h_i^*)$, set then $z_i(t)$ equal to such $h_i$. By an analogous argument, for all $t \in P_i(\tau_i(h_i^*))$ with $t_i = \tau_i(h_i^*)$, set $z(t) = h_i^*$ and let $z(t)$ be any $h_i \in \tau_{i}^{-1}$ for $t \not\in P_i(\tau_i(h_i^*))$. 

 It is clear that $\tau_i(z_i(t))=t_i$, as $z_i(\tau_i(h_i^*), t_{-i})=h_i^*$ and $z_i(t) \in \tau_i^{-1}(t_i)$ for $t_i \ne \tau_i(h_i^*)$. Moreover, the argument above implies that for all $t \in T$:
 $$
 (\eta_{i}(z(t)), \lambda_i(h_i^*, t_{-i})) \cap \sol(\gamma, t) \ne \emptyset.
 $$
This concludes the proof.
 \end{proof}

 \begin{proof}[Proof of \autoref{second_lemma}]
 Fix any $t_{-i} \in T_{-i}$, and let $SOL_i(h_i^*)=\{ h_{-i} \in H_{-i}: (h_i^*, h_{-i}) \in SOL_i\}$. As $h^* \in K_i(SOL_i)$ we have:
 $$
 \int_{SOL_i(h_i^*)} 1 \de \phi_{i}(h_i^*)[h_{-i}]=1.
 $$
 As $T_{-i}$ is finite, we can break down the sum over all type profiles of $i$'s opponents as follows:
 $$
 \sum_{t_{-i} \in T_{-i}} \int_{SOL_i(h_i^*)} \indicator_{\{\tau_{-i}(h_{-i})=t_{-i}\}} (h_{-i}) \de \phi_{i}(h_i^*)[h_{-i}]=\int_{SOL_i(h_i^*)} 1 \de \phi_{i}(h_i^*)[h_{-i}]=1
 $$
 Moreover, as $h^* \in K_i(W_i)$, it  holds for all $t_{-i} \in T_{-i}$:
 $$
 p_i(t_{-i}|\tau_i(h_i^*)) = \int_{H_{-i}} \indicator_{\{\tau_{-i}(h_{-i})=t_{-i}\}} (h_{-i}) \de \phi_{i}(h_i^*)[h_{-i}] \geq \int_{SOL_i(h_i^*)} \indicator_{\{\tau_{-i}(h_{-i})=t_{-i}\}} (h_{-i}) \de \phi_{i}(h_i^*)[h_{-i}]
 $$
 The last inequality follows from  $SOL_i(h_i^*) \subseteq H_{-i}$. As the sum over types of both sides is 1, from this inequality it follows that:
 $$
 p_i(t_{-i}|\tau_i(h_i^*)) =  \int_{SOL_i(h_i^*)} \indicator_{\{\tau_{-i}(h_{-i})=t_{-i}\}} (h_{-i}) \de \phi_{i}(h_i^*)[h_{-i}].
 $$
 Thus, whenever $p_i(t_{-i}|\tau_i(h_i^*))>0$, the right-hand side of the inequality above is strictly positive. This means there exists at least one $h_{-i}\in H_{-i}$ such that $h_{-i} \in SOL_i(h_i^*)$ and $\tau_{-i}(h_{-i})=t_{-i}$. Given that $h_{-i} \in SOL_i(h_i^*)$ implies $(h_i^*, h_{-i}) \in SOL_i$, this concludes the proof.
 \end{proof}

\end{appendices}
\end{document}

%% file: support_files/figure.tex
\begin{figure}[ht]
\centering
\begin{minipage}[c]{0.45\linewidth}
\begin{center}
    \textbf{Social Choice Functions}
\end{center}
\begin{tikzpicture}
\draw[very thick] (0,0) rectangle (\linewidth,0.75\linewidth);  
\node [at={(1.5,0.15)}] {\scriptsize All Solution Concepts};
\draw[very thick] (0.5\linewidth,0.35\linewidth) ellipse[x radius = 0.47\linewidth, 
y radius = 0.3\linewidth];
\node [at={(0.5\linewidth,0.6\linewidth)}] {\scriptsize WSC = NecBIC};
\filldraw[black] (0.4\linewidth,0.2\linewidth) circle (2pt) node[anchor=east]{\scriptsize Level-k};
\filldraw[black] (0.8\linewidth,0.3\linewidth) circle (2pt) node[anchor=east]{\scriptsize BNE};
\filldraw[black] (0.5\linewidth,0.4\linewidth) circle (2pt) node[anchor=east]{\scriptsize ICR};
\filldraw[black] (0.8\linewidth,0.7\linewidth) circle (2pt) node[anchor=east]{\scriptsize CE};
\filldraw[black] (0.2\linewidth,0.7\linewidth) circle (2pt) node[anchor=east]{\scriptsize NBE};
\end{tikzpicture}
\end{minipage}
\begin{minipage}[c]{0.45\linewidth}
\begin{center}
    \textbf{Social Choice Sets}
\end{center}
\begin{tikzpicture}
\draw[very thick] (0,0) rectangle (\linewidth,0.75\linewidth);  
\node [at={(1.5,0.15)}] {\scriptsize All Solution Concepts};
\draw[very thick] (0.5\linewidth,0.35\linewidth) ellipse[x radius = 0.47\linewidth, 
y radius = 0.3\linewidth];
\node [at={(0.5\linewidth,0.6\linewidth)}] {\scriptsize WSC};
\node [at={(0.77\linewidth,0.45\linewidth)}] {\scriptsize NecBIC};
\draw[very thick] (0.75\linewidth,0.35\linewidth) ellipse[x radius = 0.2\linewidth, 
y radius = 0.16\linewidth];
\filldraw[black] (0.4\linewidth,0.2\linewidth) circle (2pt) node[anchor=east]{\scriptsize Level-k};
\filldraw[black] (0.8\linewidth,0.3\linewidth) circle (2pt) node[anchor=east]{\scriptsize BNE};
\filldraw[black] (0.8\linewidth,0.7\linewidth) circle (2pt) node[anchor=east]{\scriptsize CE};
\filldraw[black] (0.2\linewidth,0.7\linewidth) circle (2pt) node[anchor=east]{\scriptsize NBE};
\filldraw[black] (0.5\linewidth,0.4\linewidth) circle (2pt) node[anchor=east]{\scriptsize ICR};
\end{tikzpicture}
\end{minipage}  
    \caption{\footnotesize The class of solution concepts such that BIC is necessary for implementation (NecBIC) of all SCFs coincides with the class of WSC solution concepts for full implementation of SCFs (left) and it is a subset of the class of WSC solution concepts for full implementation of SCSs (right).}
        \label{fig:diagram}
\end{figure}
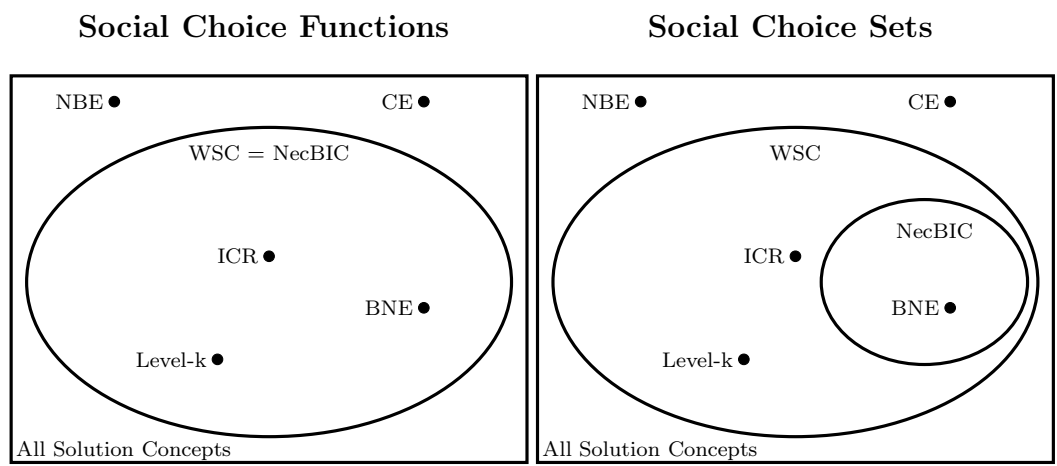